\documentclass[aps, prd, 10pt, twocolumn, superscriptaddress, noshowpacs, preprintnumbers, longbibliography, groupedaddress, footinbib, bibnotes]{revtex4-1}

\usepackage{amsmath}
\usepackage{amsfonts}
\usepackage{amssymb}
\usepackage{dsfont}
\usepackage{bbold}
\usepackage{epsfig}
\usepackage{graphicx}
\usepackage{bm}
\usepackage{array}
\usepackage{hyperref}
\usepackage{listings}
\usepackage{color}
\usepackage[normalem]{ulem}
\usepackage{cancel}

\newcommand{\vb}{\vec{v}_{\mathrm{b}}}
\newcommand{\vbr}{v_{\mathrm{b}}^{r}}
\newcommand{\den}{\rho(\theta,t)}
\newcommand{\denbar}{\bar{\rho}(\theta,t)}
\newcommand{\denp}{\rho(\theta^\prime,t)}
\newcommand{\denpbar}{\bar{\rho}(\theta^\prime,t)}

\newcommand{\denep}{J_{e}(\theta^\prime,t)}
\newcommand{\denepbar}{\bar{J}_{e}(\theta^\prime,t)}

\begin{document}

\title{Fast flavor conversion of neutrinos in presence of matter bulk velocity}

\author{Ian Padilla-Gay}
\email{ian.padilla@nbi.ku.dk}
\thanks{ORCID: \href{https://orcid.org/0000-0003-2472-3863}{0000-0003-2472-3863}}
\affiliation{Niels Bohr International Academy \& DARK, Niels Bohr Institute,\\ University of Copenhagen, Blegdamsvej 17, 2100 Copenhagen, Denmark}

\author{Shashank Shalgar}
\email{shashank.shalgar@nbi.ku.dk}
\thanks{ORCID: \href{https://orcid.org/0000-0002-2937-6525}{0000-0002-2937-6525}}
\affiliation{Niels Bohr International Academy \& DARK, Niels Bohr Institute,\\ University of Copenhagen, Blegdamsvej 17, 2100 Copenhagen, Denmark}

\date{\today}

\begin{abstract}
A dense gas of neutrinos and antineutrinos can undergo fast pairwise conversions near the decoupling regions of core-collapse supernovae and in compact binary neutron star mergers. The flavor dependent neutrino heating can play a role in sustaining convection inside the hot and dense matter. In this paper, we study the unexplored effect of the bulk velocity of matter on fast pairwise conversions and demonstrate that depending on the direction and the magnitude of the bulk velocity, neutrino flavor conversions could be significantly enhanced or suppressed. The bulk velocity of matter, which is usually neglected in the context of neutrino oscillations, can reach values of one-tenth of the speed of light in astrophysical environments. We find that bulk velocities much smaller than the maximum allowed velocities can substantially change the neutrino flavor conversion rate. The demonstration of possible enhancement of neutrino flavor conversion rate due to the bulk velocity of matter also raises several important issues that are of relevance in the supernova mechanism. Future studies with realistic velocity profiles could elucidate the possible implications on the phenomenology of core-collapse supernovae and neutron star mergers.
\end{abstract}

\maketitle

\section{Introduction}\label{sec:intro}

Core collapse supernovae are one of the densest astrophysical objects which copiously produce neutrinos. 
Three flavors of neutrinos and antineutrinos are produced within about 10 seconds which carry most of the energy produced in the supernova. The neutrinos produced within this short duration play a vital in the supernova explosion mechanism according to the
delayed neutrino-driven supernova explosion mechanism~\cite{Bethe:1984ux}. 
The neutrino-driven explosion mechanism is often invoked to explain the revival of the shock after it loses its energy photo-dissociating iron group nuclei in the outer parts of the core. The stalled shock is revived because of the nonuniform deposition of energy by neutrinos in the outer envelope of the supernova, which leads to convection, resulting in convection driven hydrodynamical instabilities~\cite{Janka:2016fox}.

The energy deposition by neutrinos is flavor dependent as the electron type neutrinos have a larger cross section with matter than non-electron type neutrinos. The presence of flavor conversion at the right time and radius can thus affect the energy deposition by neutrinos and their role in the supernova mechanism. Neutrino flavor conversions can play a role in modifying the neutrino-driven explosion mechanism only if the flavor conversions can occur relatively early and at small radii where the densities are extremely large. 

However, the computation of neutrino flavor evolution in dense astrophysical environments such as supernovae is an extremely challenging task. Due to the high number density of neutrinos, the neutrino flavor evolution is nonlinear as a consequence of the coherent forward scattering of neutrinos from other neutrinos~\cite{Pantaleone:1992eq}, also known as neutrino self-interactions. This phenomenon is similar to the well-known MSW effect~\cite{Mikheev:1986gs, Wolfenstein:1977ue}; however, in the case of neutrino self-interactions, the equation of motion that governs the flavor evolution is nonlinear, and this leads to a rich and interesting phenomenology.

One of the first numerical implementations of the neutrino self-interactions in supernovae made several simplifying assumptions, like perfect spherical symmetry and instantaneous decoupling of all neutrinos at the same radius~\cite{Duan:2006jv, Duan:2006an, Duan:2005cp, Duan:2010bg}. The simplifying assumptions made in the earlier calculations were found to be unjustified for several reasons. The nonlinear evolution of neutrino flavor evolution can lead to spontaneous breaking of symmetries initially present in the system~\cite{Raffelt:2013rqa, Duan:2014gfa, Abbar:2015mca}, due to which flavor transformations can occur in much denser environments. The numerical simulations carried out in the simplified setup, however, demonstrated conclusively that in the presence of neutrino self-interactions, neutrinos with different momenta evolve in a `collective' manner. The term `collective neutrino oscillations' is often used to refer to neutrino flavor evolution in the presence of neutrino self-interactions as a result.

Also, the non-instantaneous decoupling of neutrinos can lead to different angular distributions with different flavors, and in some cases, lead to angular distributions that have a crossing in the electron lepton number (ELN)~\cite{Shalgar:2019kzy}. The presence of a crossing in the ELN is a necessary and sufficient condition for the general case of an inhomogeneous neutrino ensemble~\cite{Morinaga:2021vmc}; however in the homogenous case, which we assume throughout this work, only a special class of ELN crossings lead to flavor instabilities. In other words, the ELN crossing by itself is not a sufficient condition if spatial  homogeneity is assumed.

Unlike the slow collective neutrino flavor evolution studied earlier~\cite{Duan:2006jv, Duan:2006an, Duan:2005cp, Duan:2010bg}, which can occur only in environments for which the neutrino self-interaction potential is comparable to the vacuum frequency, the fast flavor conversions can occur in much larger densities. 

The large density of neutrinos in the vicinity of the proto-neutron star (PNS) may foster fast conversions~\cite{Tamborra:2020cul}. 
The timescale on which fast pairwise conversions occur is determined by the local (anti)neutrino number density. For typical values of the neutrino densities near the decoupling region, fast oscillations could take place on a scale as small as a few nanoseconds (i.e., a few meters)~\cite{Capozzi:2019lso, Martin:2019gxb, Johns:2019izj, Johns:2020qsk, Dasgupta:2018ulw}, with the possibility of pair conversions taking place in the deep interior of supernovae~\cite{Glas:2019ijo}. Although there has been a serious discussion regarding the role neutrino flavor conversions may play in instigating convection~\cite{Pejcha:2011en, Suwa:2011ac, Pllumbi:2014saa}, in this paper, we focus on the converse; the effect of convection on neutrino flavor evolution. It should be noted that in the past, there have been studies of nontrivial matter profiles on collective neutrino flavor evolution, but they have been mostly limited to the study of the possible effect of turbulence and the small scale spatial variations that can induce parametric resonances~\cite{Friedland:2006ta, Kneller:2010sc, Choubey:2007ga, Akhmedov:1999ty, Kneller:2012id,Ma:2018key}.

We study the effect of the bulk velocity of matter on fast pairwise conversions. Depending on the direction of the bulk velocity and the magnitude, neutrino flavor conversions may get enhanced or suppressed. In the convective region of supernova, the bulk velocity of matter can, therefore, can lead to enhancement in some regions and suppression in other regions. This can lead to a differential in the neutrino heating rate, which could feed the convection. 

It should be noted, however, that the existence of large number densities of neutrinos is not restricted to the interior of supernovae. Large number densities of neutrinos are produced in the immediate aftermath of neutron star mergers, where favorable conditions for the development of fast flavor conversions may exist~\cite{Wu:2017qpc,Wu:2017drk}. Although the neutrino flavor evolution is not expected to play a role in the dynamics of the neutron star mergers, the synthesis of heavy elements by $r$-process can be affected by neutrino flavor evolution~\cite{George:2020veu, Li:2021vqj}. A detailed study on the flavor evolution of neutrinos in compact binary objects~\cite{Padilla-Gay:2020uxa} demonstrated that even in the presence of favorable conditions, the mixing of neutrinos may be minimal, suggesting that the assumption of flavor equipartition may overestimate the flavor conversion rate. Also, the large bulk velocities of matter are present in the region where neutrino flavor evolution occurs~\cite{Frensel:2016fge, Hotokezaka:2012ze, Kyutoku:2013wxa}. However, the geometry of neutron star mergers is typically more complicated than that of a supernova. Although the results obtained in this paper would have significant implications for neutrino flavor evolution in the context of neutron star mergers, we do not comment on the relevant phenomenological implications. 

Several complications can arise even in the case of a supernova, which makes a realistic estimate of the role of the bulk velocity extremely challenging. 
In this paper, we thus focus on a simplified case of quantifying the effect of the bulk velocity of matter in some representative setups involving a homogeneous neutrino gas.

The paper is organized as follows: In Sec.~\ref{sec:qke}, we introduce the equations of motion for neutrino flavor evolution in a medium that is not isotropic in nature. The source of the anisotropy of the medium is discussed and derived from the bulk velocity in Sec.~\ref{sec:anisotropic}. In Sec.~\ref{sec4}, we demonstrate the enhancement of neutrino flavor transformation due to the presence of the bulk velocity. In Section~\ref{linearstability}, we demonstrate the agreement between our numerical simulations and the semianalytical technique of linear stability analysis. Section~\ref{sec:grid} presents the main results of this paper in the form of showing the effect of bulk velocity on the evolution of the neutrino flavor evolution by considering the evolution of a family of angular distributions parameterized in a simple manner. Finally, in Sec.~\ref{sec:conclusion} we conclude and offer an outlook.


\section{Neutrino equations of motion in dense media}
\label{sec:qke}

This section describes the equations of motion governing the flavor evolution of neutrinos in the presence of a neutrino and electron background. Among other simplifications, we assume a two flavor system where $\nu_{e}$ is the electron flavor and $\nu_{x}$ represents a mixture of the $\nu_{\mu}$ and $\nu_{\tau}$ flavors. 
The flavor content of the neutrinos and antineutrinos can, therefore, be encoded by a $2\times 2$ density matrix, which we denote by $\rho$ and $\bar{\rho}$, respectively. Although, the two flavor approximation has its limitations in the context of fast flavor conversions~\cite{Chakraborty:2019wxe, Capozzi:2020kge, Shalgar:2021wlj}, the formalism used in this paper can be extended to the three flavor case, and the central premise of the results remain unchanged. It should be noted that the three flavor effect acts sequentially for the slow collective phenomenon~\cite{Dasgupta:2010cd, Dasgupta:2010ae, Dasgupta:2007ws, Friedland:2010sc}, but that is not the case for fast flavor conversions.

Fast flavor conversions can be expected to occur in dense regions of supernovae where the effect of collisions cannot be ignored. For simplicity, we also ignore the effect of momentum changing collisions, which can dramatically enhance the flavor conversions rates depending on the initial conditions used in the simulations~\cite{Shalgar:2020wcx}.

We also assume that the angular distribution of neutrino has azimuthal symmetry with respect to the radial direction, even though the azimuthal symmetry breaking effects in the case of fast flavor conversion can have interesting consequences~\cite{Shalgar:2021oko}. Because of this simplifying assumption, the angular distribution of neutrinos can be expressed in terms of a single polar angle denoted by $\theta$. We also ignore the effect of spatial inhomogeneity that can arise in collective neutrino oscillations so that the evolution neutrino flavor only depends on the polar angle, $\theta$, and time, $t$. Furthermore, for simplicity, we assume that (anti)neutrinos are of a single energy.

\begin{eqnarray}
\den = 
\begin{pmatrix}
\rho_{ee} & \rho_{ex}\\
\rho_{ex}^{*} & \rho_{xx} 
\end{pmatrix}\ \quad\ \mathrm{and}\ \quad\ 
\denbar =
\begin{pmatrix}
\bar{\rho}_{ee} & \bar{\rho}_{ex}\\
\bar{\rho}_{ex}^{*} & \bar{\rho}_{xx} 
\end{pmatrix}\ ,
\end{eqnarray}
where the diagonal terms are related to the occupation number of a given flavor of (anti)neutrinos while the off-diagonal terms describe the coherence between flavors. The equations of motion for a homogeneous neutrino gas can in general be written as the Heisenberg equations,
\begin{eqnarray}
i\frac{\partial}{\partial t}  \den
= [H(\theta),\den]\ ,
\label{eom1} \\
i\frac{\partial}{\partial t}  \denbar
= [\bar{H}(\theta),\denbar]\ .
\label{eom2}
\end{eqnarray}
In the case of a homogeneous neutrino gas, the total and the partial derivative can be used interchangeably on the left-hand side. However, if the assumption of homogeneity is relaxed, the left-hand side needs to be replaced by a total derivative consisting of a time derivative and an advective term. For an inhomogeneous initial condition, the advective term can modify the neutrino flavor evolution, but we ignore this effect here~\cite{Shalgar:2019qwg}.

The Hamiltonian receives the contributions from neutrino vacuum oscillations, electron-neutrino coherent forward scattering and neutrino-neutrino coherent forward scattering,
\begin{eqnarray}
\label{eq:H}
H(\theta) = H_{\mathrm{vac}} +  H_{e\nu}(\theta)+H_{\nu\nu}(\theta) \ .
\end{eqnarray}

The vacuum term is proportional to the characteristic vacuum frequency $\omega = \Delta m^2 / 2 E$,
\begin{eqnarray}
	H_{\mathrm{vac}} = \frac{\omega}{2} \left(\begin{matrix}-\cos 2\vartheta_V & \sin 2\vartheta_V\\ \sin 2\vartheta_V & \cos 2\vartheta_V\end{matrix}\right).
\end{eqnarray}
Here, we assume that all (anti)neutrinos have the same energy $E$ to mimic the evolution of neutrinos with an energy distribution peaking at the value $E$~\cite{Shalgar:2020xns}. For antineutrinos we can obtain $ \bar{H}$ by replacing $ \omega \rightarrow -\omega$ in $H_{\mathrm{vac}}$.

The flavor transition probability is modified due to the presence of electrons in the medium. The modification of neutrino flavor conversion probability depends on the electron number density and the direction of motion. In the absence of a bulk velocity in the medium, the direction dependence of the matter effect averages out to zero. It is thus common to ignore the angle dependence of the matter term in the Hamiltonian. 
However, ignoring the angle dependence is an approximation and the proper expression for the matter term in the most general case is,  
\begin{eqnarray}
\label{eq:Hmat_vdotv}
H_{e\nu}(\vec{v}) &=& 
 \lambda \int d\vec{v}_{e} \left[\denep - \denepbar \right] \left[1 - \vec{v} \cdot\vec{v}_{e} \right] \ ,
\end{eqnarray}
where $\lambda \equiv \sqrt{2} G_{\textrm{F}} n_{e}$ parametrizes the strength of matter potential, which is proportional to the Fermi constant, $G_{\textrm{F}}$ and the number density of electrons, $n_{e}$. $J_{e}(\theta^{\prime})$ and $\bar{J_{e}}(\theta^{\prime})$ describes the angular distribution of the momentum of electrons and positrons, respectively, and normalized such that the $\int_{-1}^{1}J_{e}(\theta^{\prime}) d\cos\theta^{\prime}=1$. We ignore the postiron density hereafter. Moreover, $\vec{v}$ and $\vec{v}_{e}$ are the velocities of the test neutrino and the matter background, respectively. For a sufficiently homogeneous system, the momentum or angular distribution of electrons can be completely described by the bulk velocity denoted by $\vb$. 

The standard matter effect considered in the literature is obtained by setting $\vec{v}_{e}$ equal to zero. 
However, in the interior of a supernova, the matter can flow can occur near the neutrino decoupling region either due to the matter infall onto the proto-neutron star or due to plumes of heated matter rising radially outwards. The velocity in question is much smaller in magnitude than the speed of light but not small enough to be negligible.

The magnitude of the bulk velocity in the vicinity of the neutrino decoupling region can be as large as  $|\vb| \simeq 0.1$~\cite{Just:2018djz}. We assume that neutrinos travel at the speed of light, $|\vec{v}| = 1$, where all velocities are expressed in units of speed of light, unless otherwise specified. Therefore, $H_{e\nu}$ can be written as,
\begin{eqnarray}
\label{eq:Hmat}
H_{e\nu}(\theta) &=& \lambda \int d (\cos{\theta^\prime}) \denep  \left[1 - |\vb| \cos{\theta}\cos{\theta^{\prime}} \right]\ .
\end{eqnarray}
Notice that assuming an anisotropic momentum distribution of electrons i.e. their bulk velocity is not zero, introduces a new contribution to the equations of motion which is captured by Eq.~\ref{eq:Hmat}. The angular dependence of $J_{e}(\theta^{\prime}) d(\cos{\theta^{\prime}})$ is determined by $\vb$ alone as described in section~\ref{sec:anisotropic}.

The third term in the Hamiltonian is the neutrino self-interaction term which describes the potential experienced by neutrinos due to other neutrinos in the medium,
\begin{eqnarray}
\label{eq:Hnunu}
H_{\nu\nu}(\theta) &=& 
\mu \int d (\cos{\theta^\prime}) \left[\denp - \denpbar \right] \left[1 - \cos{\theta}\cos{\theta^{\prime}} \right] \ , \nonumber \\
& & 
\end{eqnarray}
where $\mu$ parametrizes the interaction strength of neutrinos among themselves and is proportional to the number density of (anti)neutrinos. The velocity dependence for the neutrino self-interaction potential is the same as in the matter potential due to the vector-vector coupling of weak interactions in both cases.

We focus on the evolution of the neutrino flavor due to the matter term and self-interaction term, and set $\omega=0$. We instead provide a seed to the off-diagonal components of the density matrices to ensure that our results do not depend on the vacuum Hamiltonian. 
This ensures that in the absence of the bulk velocity, the neutrino flavor evolution is bipolar in nature and thus makes a comparison between different cases easier to illustrate. 
Since we are interested in the interplay between $H_{e\nu}$ and $H_{\nu\nu}$ near the PNS we assume that the neutrino self-interaction potential and the electron-neutrino interaction strength are equal to each other i.e. $\mu=\lambda=10^5 \ \mathrm{km}^{-1}$, which is representative of the conditions realizable in a supernova in the decoupling region. It becomes more transparent that the flavor instability condition, mainly determined by crossing between $\rho_{ee}$ and $\bar{\rho}_{ee}$, is modified by a new contribution which is proportional to the ratio $\lambda/\mu$ and the particular shape of $J_{e}(\theta)d(\cos{\theta})$, which can be derived from the value of the bulk velocity.

\begin{figure*}[t!]
\centering
\includegraphics[width=0.99\textwidth]{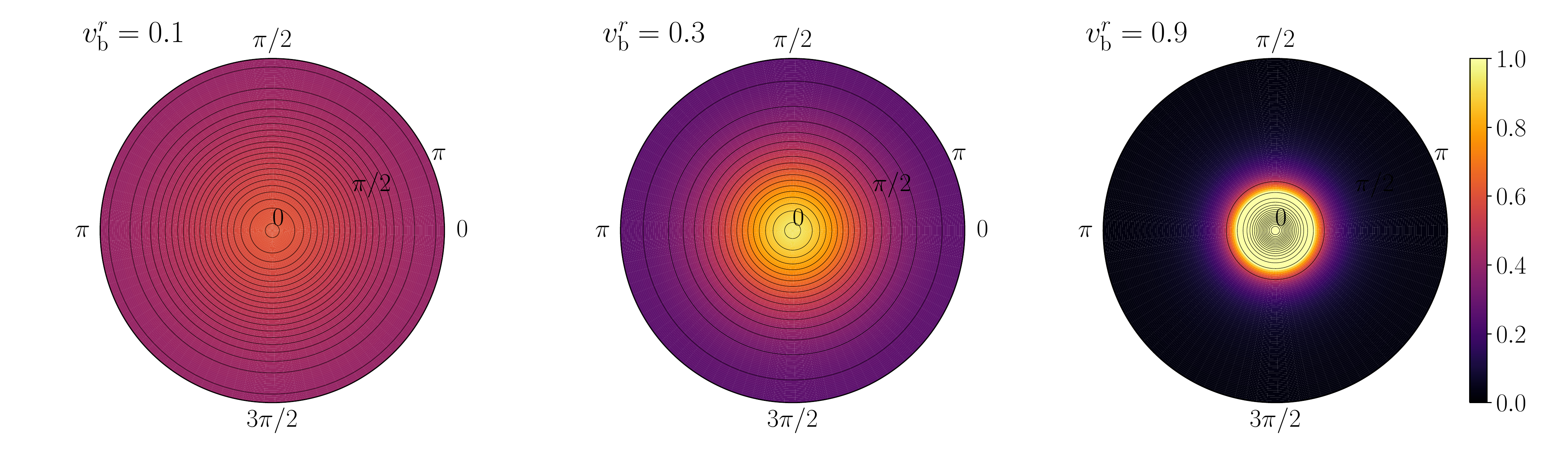}
\caption{
Two dimensional Lambert projections of the function $ {d\widetilde{\cos{\theta}}}/{ d\cos{\theta} } $ as shown in Eq.~\ref{eq:ne_aniso} for $\vbr=0.1$ (left), $\vbr=0.3$ (middle) and $\vbr=0.9$ (right). The solid lines are the isocontours of constant matter distribution to show the symmetry along the azimuthal direction. A non-vanishing $\vbr$ leads to the effect of relativistic beaming, which is clearly visible for $\vbr=0.9$.
}
\label{fig:1pr}
\end{figure*}


\section{Matter anisotropy}
\label{sec:anisotropic}

This section describes how the angular distribution of electrons and the bulk velocity are related to each other. 
In the absence of bulk velocity, which is generally considered in the literature, the matter term is independent of direction. Upon the introduction of a non-vanishing bulk velocity, the matter term can be thought of as the Lorentz boosted version of the usual matter term. We start with an angular distribution for electrons that would be isotropic for an observer traveling along with the fluid element. We denote this reference frame by $S$. On the other hand, we also consider another reference frame denoted by $\widetilde{S}$ which is fixed with respect to the center of the supernova, and in which the matter travels with a velocity $\vb$. 
Throughout this paper the bulk velocity is assumed to be in either radially outwards or inwards direction. The radial velocity of matter, denoted by $\vbr$, is always the same in magnitude as $\vb$, but $\vbr$ can be negative for bulk velocities that are radially inward. Obtaining the angular distribution of electrons in $\widetilde{S}$ can be done in a straightforward manner, as we show in this section, and provides us with a unique angular distribution for electrons once the bulk velocity is specified as long as the electrons are highly relativistic. 

The starting point is to consider an isotropic distribution of electrons in reference frame $S$.
The same isotropic distribution will be peaked along the direction of propagation for another observer in reference frame $\widetilde{S}$. 
Equivalently, we can calculate the angular distribution of electron momenta for a given bulk velocity $\vb$ by boosting a relativistic isotropic gas. We denote the angular distribution in reference frame $S$ by $f_{e}$, which by definition is independent of $\theta$.

\begin{figure*}[t!]
\centering
\includegraphics[width=0.95\textwidth]{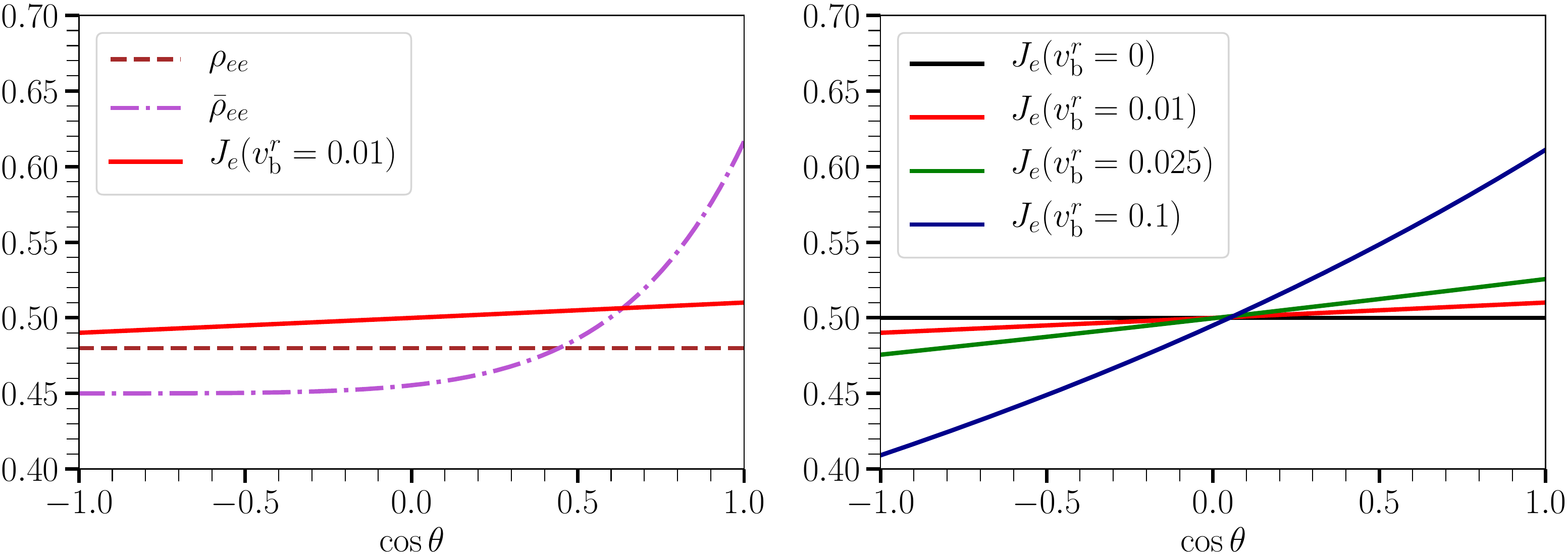}
\caption{ 
\textit{Left:} (Anti)neutrino and matter angular distributions according to Eq.~\ref{eq:init} with $\delta=-0.02$, $\sigma_{\nu}=0.6$ and $\vbr=0.01$. \textit{Right:} Various angular distributions of matter as a function of $\vbr$; as the magnitude of $\vbr$ increases the greater is the difference between the forward ($\cos{\theta}=1$) and the backward ($\cos{\theta}=-1$) directions.
}
\label{fig:2pr}
\end{figure*}

The angular distribution for electrons achieved after boosting $f_{e}$ with a given bulk velocity serves as a good approximation for the angular distribution that can be expected in a realistic astrophysical system. In order to understand how the anisotropy arises from the inclusion of the bulk velocity, we define the parallel and perpendicular components of the velocity of electrons $\vec{v}_{e}$ (with respect to the bulk velocity) as $v_{e}^{||}$ and $v_{e}^{\bot}$, respectively. 
In the limit of vanishing bulk velocity, the velocity of electrons $\vec{v}_{e}$ is isotropically distributed, and the magnitude is determined by the temperature scale. In the interior of a supernova, the temperatures are typically much larger than the mass of the electron, and hence $|\vec{v}_{e}|$ almost equal to the speed of light.
The Lorentz transformed quantities are denoted by a tilde i.e. $\widetilde{v}_{e}^{||}$ and $\widetilde{v}_{e}^{\bot}$. Assuming that the bulk velocity oriented along the parallel direction i.e. $\vb \propto v_{e}^{||}$, then, relativistic addition of velocities leads to the following relations, 
\begin{eqnarray}
\label{eq:ve}
	\widetilde{v}_{e}^{||}=\frac{{v}_{e}^{||} - \vbr}{1 - {v}_{e}^{||} \vbr}, \ \quad \mathrm{and} \ \quad \widetilde{v}_{e}^{\bot}=\frac{{v}_{e}^{\bot}}{\Gamma(1 - {v}_{e}^{\bot} \vbr )} \ , 
\end{eqnarray}
where $\Gamma=1/\sqrt{1-[\vbr]^2}$ and $\vbr$ is the radial component of the bulk velocity as defined earlier in this section. Due to the velocity transformation the relative angle between electrons changes as a function of the direction of the bulk velocity $\vb$ according to 
\begin{eqnarray}
\label{eq:tan}
	\widetilde{\tan{\theta}} = \frac{ \widetilde{v}_{e}^{\bot} }{ \widetilde{v}_{e}^{||} } = \frac{ {v}_{e}^{\bot} }{\Gamma ({v}_{e}^{||} - \vbr) } = \frac{ {v}_{e}\sin{\theta} }{ \Gamma({v}_{e}\cos{\theta} - \vbr) }\ . 
\end{eqnarray}
Moreover, by substituting the definition of the boosted (unboosted) parallel components $\widetilde{v}_{e}^{||}=\widetilde{v}_{e}\widetilde{\cos{\theta}}$  ($v_{e}^{||}=v_{e}\cos{\theta}$) in Eq.~\ref{eq:ve} and assuming that electrons are highly relativistic i.e. $\widetilde{v}_{e} \simeq v_{e} \simeq 1$ we obtain a similar relation to Eq.~\ref{eq:tan} in terms of the cosine of the angle,
\begin{eqnarray}
\label{eq:aberr}
	\widetilde{\cos{\theta}} = \frac{ \cos{\theta}-\vbr }{ 1 - \vbr\cos{\theta} }\ .
\end{eqnarray}
Equation \ref{eq:aberr} can be used to derive the angular distribution of matter. The flux of electrons between $\cos\theta$ and $\cos\theta + d \cos\theta$ is given by $f_{e}(\theta) d\cos\theta$ in the unboosted frame. The same quantity in the boosted frame given by $\widetilde{f_{e}}(\theta) d\widetilde{\cos{\theta}}$ is related to the one in the unboosted frame by the following formula,
\begin{widetext}
\begin{eqnarray}
\label{eq:ne_aniso}
\widetilde{f_{e}}(\theta) d\widetilde{\cos{\theta}} = \widetilde{f_{e}}(\theta) \frac{d\widetilde{\cos{\theta}}}{d\cos{\theta}} d\cos{\theta} =f_{e}(\theta) \Bigg( \frac{1-[\vbr]^2}{(1-\vbr\cos{\theta})^2} \Bigg) d\cos{\theta}\ ,
\end{eqnarray}
\end{widetext}
where we have inserted the derivative of $\widetilde{\cos{\theta}}$ with respect to the unboosted variable $\cos{\theta}$. Moreover, in the last step, we have assumed that $ \widetilde{f_{e}} = f_{e} $ because the contribution from this term to the anisotropy is a subleading contribution due to the highly relativistic velocities of individual electrons. The anisotropy of the matter potential thus generated is illustrated in Fig.~\ref{fig:1pr}. As the bulk velocity increases, the matter potential becomes more and more peaked. Also, we can see that the matter potential is azimuthally symmetric around the direction of the bulk velocity as expected. 

\begin{figure*}[t!]
\centering
\includegraphics[width=0.99\textwidth]{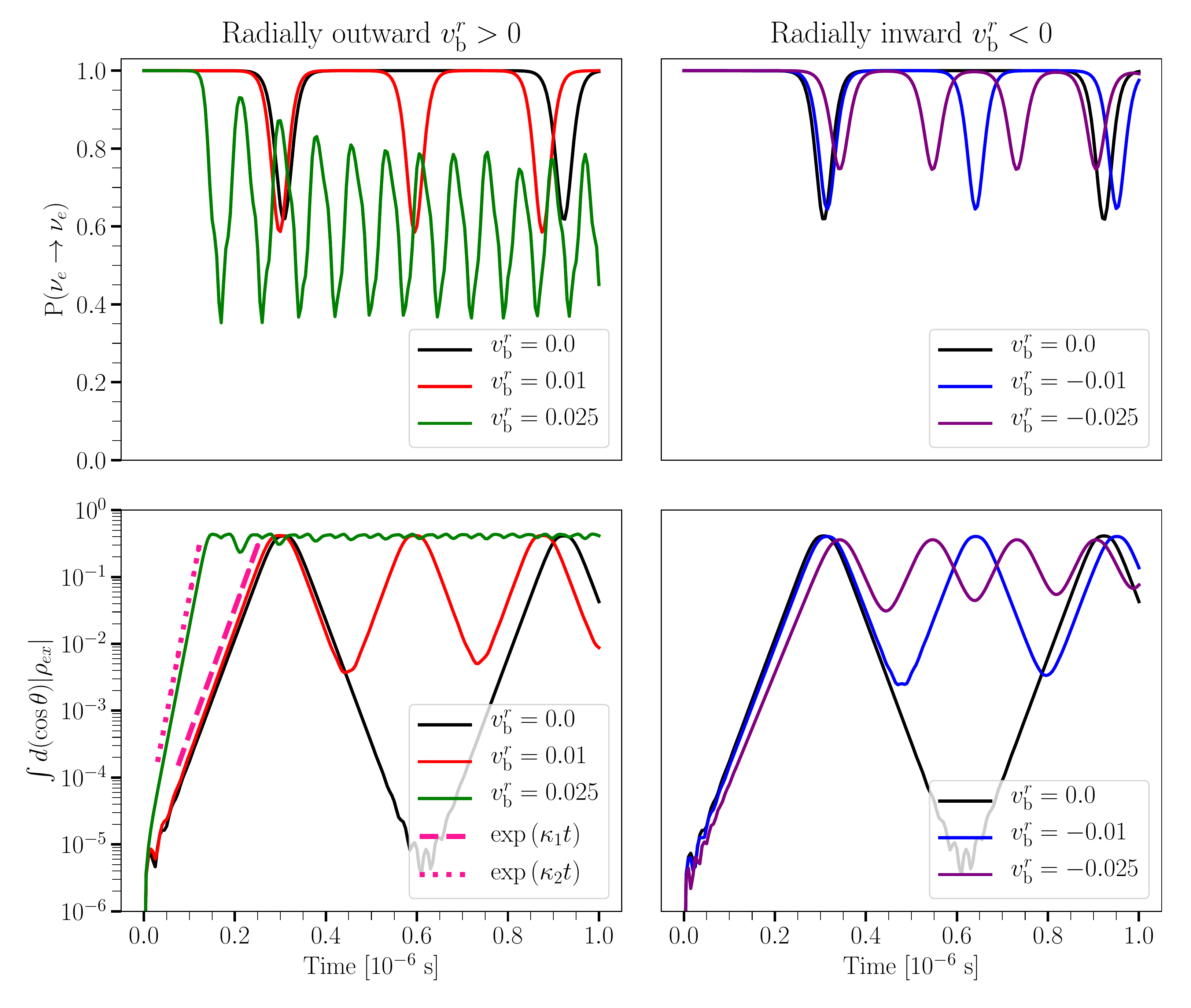}
\caption{Survival probability $\mathrm{P}(\nu_e \rightarrow \nu_e)$ (top panels) and the angle-integrated modulus of the off-diagonal term $\rho_{ex}$ (bottom panels) for the scenarios where matter moves radially outward (left panels) and radially inward (right panels). The role of $\vbr \neq 0$ is twofold: Firstly, there is a enhancement of conversions for $\vbr>0$ compared to the $\vbr \leq 0$ cases, see top panels. Secondly, oscillations set in faster i.e., $\vbr > 0$ leads to larger growth rates such as $\kappa_1/\mu=0.0027$ (dotted line) and $\kappa_2/\mu=0.0014$ (dashed line), as shown in the bottom panels.
}
\label{fig:3pr}
\end{figure*}


\section{Flavor evolution for anisotropic matter}
\label{sec4}

The angle dependent matter Hamiltonian derived in the previous section can be used in the numerical simulation of fast flavor evolution to uncover the effect of bulk velocity of matter. In this section, we specify a class of angular distributions for (anti)neutrinos that are characterized by the parameters $\sigma_{\nu}$ and $\delta$, which control the width of $\bar{\rho}_{ee}$ and the relative normalization between $\rho_{ee}$ and $\bar{\rho}_{ee}$, respectively:
\begin{eqnarray}
\label{eq:init}
	\rho_{ee}(\theta) &=& 0.5 \ , \nonumber \\
	\bar{\rho}_{ee}(\theta) &=&  0.45 -\delta + \frac{0.1}{\sigma_{\nu}} \exp{\Bigg(\frac{-\theta^2}{2\sigma_{\nu}^2}\Bigg)} \ , \nonumber \\
	J_e(\theta) &=& 0.5 \frac{1-[\vbr]^2}{(1-\vbr\cos{\theta})^2}  \ .
\end{eqnarray}


We implement the angular distribution for (anti)neutrinos and matter presented in Eq.~\ref{eq:init} and focus on the the neutrino flavor evolution as $J_{e}$ transitions from the isotropic ($\vbr=0$) to the anisotropic case ($\vbr \neq 0$). Fig.~\ref{fig:2pr} shows the effect of radially outwards velocity on the angular distribution of the matter term.

In Fig.~\ref{fig:3pr}, we show the impact of the bulk velocity on a representative angular distribution of (anti)neutrinos (shown in Fig.~\ref{fig:2pr}), for various values of the bulk velocity. The top panel shows the modification due to the bulk velocity that is in the radially outward direction ($\vbr>0$), while the bottom panel shows the modification due to a bulk velocity in the radially inward direction ($\vbr<0$). Figure~\ref{fig:3pr} shows that while it is possible to get an enhancement in flavor conversion probability due to a bulk velocity that is radially outward, the radially inward bulk velocity tends to suppress the flavor conversion. The lower panels of Fig.~\ref{fig:3pr} show that the effect of the bulk velocity on the flavor conversion probability is present in the linear regime of the evolution that is characterized by $\rho_{ex} \ll (\rho_{ee}+\rho_{xx})$. In the linear regime, the nonlinear equation of motion for the flavor evolution can be reliably approximated by linear equations of motion that can be solved semi-analytically. We perform linear stability analysis, which is characterized by an exponential growth of the off-diagonal terms of the density matrix, which serves two purposes. Firstly, the matching growth rates with our numerical simulations support the validity of our numerical simulations. Secondly, the linear stability analysis demonstrates that introducing the bulk velocity for matter in the system changes the flavor stability criteria and does not just change the magnitude \ of flavor conversion in the nonlinear regime. The formalism for linear stability analysis used to present the results in the lower panels of Fig.~\ref{fig:3pr} is presented in the Sec.~\ref{linearstability}. 

Figure~\ref{fig:3pr} shows that favorable conditions for fast flavor depend on the interplay between $H_{\nu\nu}$ and $H_{e\nu}$ which is visible in the survival probability $\mathrm{P}(\nu_{e}\rightarrow \nu_{e})$, and that is that significant neutrino conversions are not achieved for arbitrarily large values of the bulk velocity $\vbr$ but are only possible when the anisotropy in the matter term is comparable with that in the neutrino-neutrino self-interaction term. In other words, a large amount of flavor conversions are possible even for modest values of $\vbr$ that may be realizable in a realistic supernova environment.


\section{Linear stability analysis}
\label{linearstability}

We validate our numerical runs with semi-analytical estimates using linear stability analysis~\cite{Banerjee:2011fj, Raffelt:2013rqa, Abbar:2015mca, Duan:2014gfa, Sawyer:2008zs}. Since we are interested in fast pairwise conversions, we study the development of flavor instabilities in the linear regime and choose the scenario where $\omega=0$ in order to focus strictly on the fast pairwise conversions i.e. $\mu \gg \omega$~\cite{Izaguirre:2016gsx}.

We start by linearizing the EoM and tracking the evolution of the off-diagonal term
\begin{eqnarray}\label{eq:ansatz}
 \rho_{ex}(\theta) = Q(\theta)e^{-i\Omega t} \ \mathrm{and}\ 
 \bar{\rho}_{ex}(\theta) = \bar{Q}(\theta)e^{-i\Omega t} ,
\end{eqnarray}
where $\Omega = \gamma + i \kappa$ represents the collective oscillation frequency for neutrinos and antineutrinos. If Im$(\Omega) \neq 0$, then the flavor instability grows exponentially with rate $|$Im$(\Omega)|$. We look for temporal instabilities for the homogeneous mode.

\begin{figure}[t!]
\centering
\includegraphics[width=0.90\columnwidth]{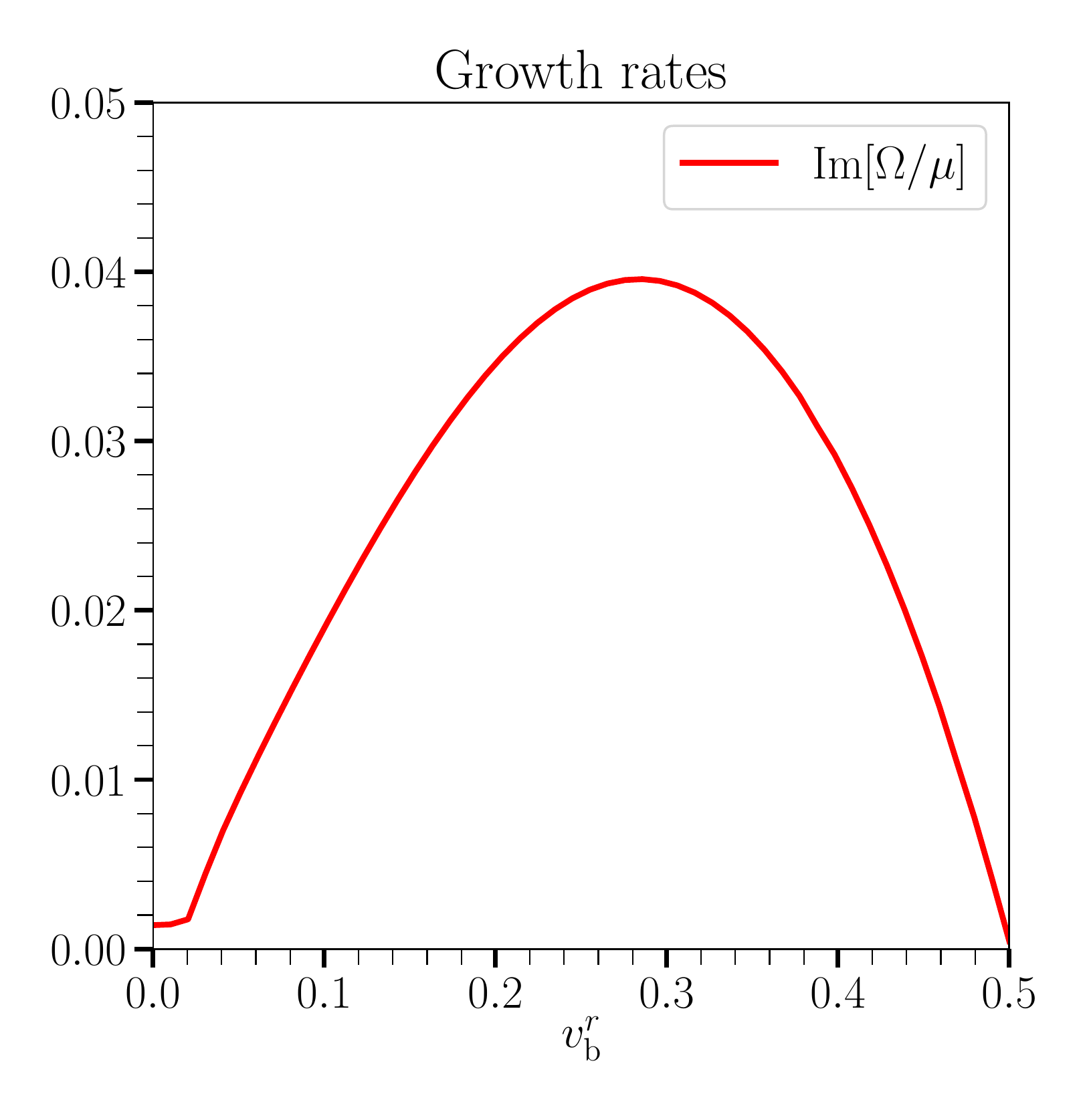}
\caption{
Magnitude of the growth rate $\mathrm{Im}[\Omega/\mu]$ as a function of $\vbr$. The angular distributions of (anti)neutrinos are fixed by choosing  $\delta=-0.02$ and $\sigma_{\nu}=0.6$ as specified in Eq.~\ref{eq:init}, while the magnitude of $\vbr$ is varied. For values $\vbr \gtrsim 0.025$ the growth rate increases with the bulk velocity. This continues until $\vbr \sim 0.3$ after which the trend is seen to reverse.
}
\label{fig:4pr}
\end{figure}

The off-diagonal component of Eq.~\ref{eom1} is 
\begin{eqnarray}\label{eom_lin}
 i \frac{\partial}{\partial t}\rho_{ex}(\theta) &=& H_{ee}(\theta)\rho_{ex}(\theta) + H_{ex}(\theta)\rho_{xx}(\theta) \nonumber \\
 &-& \rho_{ee}(\theta)H_{ex}(\theta) - \rho_{ex}(\theta)H_{xx}(\theta) \nonumber \\ 
 &=& H_{ee}(\theta)\rho_{ex}(\theta) - \rho_{ee}(\theta)H_{ex}(\theta)\ ,
\end{eqnarray}
where we have assumed $\rho_{xx} (t=0~\mathrm{s}) = \bar\rho_{xx}(t=0~\mathrm{s}) = 0$. By substituting Eq.~\ref{eq:ansatz} in the equation above and solving for $Q(\theta)$, we obtain 
\begin{eqnarray}\label{eq:Qtheta}
 Q(\theta) &=& \frac{ \rho_{ee}(\theta) \int d (\cos{\theta^\prime}) [ Q(\theta^{\prime})-\bar{Q}(\theta^{\prime}) ] \left[1 - \cos{\theta}\cos{\theta^{\prime}} \right] } { -\frac{\Omega}{\mu} + A(\theta) }\ , \nonumber \\
 & &
\end{eqnarray}
where we have defined the angle-dependent quantity $A(\theta)$ as
\begin{eqnarray}
	& &A(\theta) =  \int d (\cos{\theta^\prime}) [ \rho_{ee}(\theta^{\prime})-\bar{\rho}_{ee}(\theta^{\prime}) + \frac{\lambda}{\mu} J_e(\theta^{\prime})] \nonumber \\
	&-& \int d (\cos{\theta^\prime}) \cos{\theta}\cos{\theta^{\prime}} [ \rho_{ee}(\theta^{\prime})-\bar{\rho}_{ee}(\theta^{\prime}) + \vbr \frac{\lambda}{\mu} J_e(\theta^{\prime})]\ .  \nonumber \\
	& & \nonumber 
\end{eqnarray}
A similar procedure follows for $\bar{Q}_{\theta}$ (see Eqs.~\ref{eom2} and \ref{eq:ansatz}). Then, from combining the expressions for $Q(\theta)$ and $\bar{Q}(\theta)$, it must be true that 
\begin{eqnarray}\label{eq:QmQbar2}
 Q(\theta)-\bar{Q}(\theta) = \left[\frac{\rho_{ee}(\theta)-\bar{\rho}_{ee}(\theta)}{-\frac{\Omega}{\mu} + A(\theta)}\right] (a - b\cos{\theta})\ ,
\end{eqnarray}
where $a$ and $b$ are unknown coefficients. By substituting Eq.~\ref{eq:QmQbar2} in Eq.~\ref{eq:Qtheta}, we obtain a system of equations for the coefficients $a$ and $b$. 
\begin{eqnarray}\label{eq:system}
\begin{bmatrix}
a \\
b \\
\end{bmatrix}
=
\begin{bmatrix}
\mathcal{I}[1] & - \mathcal{I}[\cos{\theta}]  \\
\mathcal{I}[\cos{\theta}] & - \mathcal{I}[\cos{\theta}^2]  \\
\end{bmatrix}
\begin{bmatrix}
a \\
b \\
\end{bmatrix}
=
\mathrm{M}
\begin{bmatrix}
a \\
b \\
\end{bmatrix} \ ,
\end{eqnarray}
where the functional $\mathcal{I}[f]$ is
\begin{eqnarray}
 \mathcal{I}[f] = \int d(\cos{\theta}) \left[\frac{\rho_{ee}(\theta)-\bar{\rho}_{ee}(\theta)}{-\frac{\Omega}{\mu} + A(\theta)}\right] f(\theta)\ .
\end{eqnarray}
The system of equations has a not trivial solution if and only if the following condition is met
\begin{eqnarray}\label{eqn:det}
 \mathrm{det}( \mathrm{M} - \mathds{1}_{2\times2} ) = 0\ . 
\end{eqnarray}
Equation~\ref{eqn:det} is a polynomial equation in the frequency $\Omega$. To search for instabilities, we need to look for the solutions with $\mathrm{Im}(\Omega) = \kappa \neq 0$. We find the roots of this polynomial equation using the SciPy module~\cite{SciPy} in Python.

\begin{figure*}[t!]
\centering
\includegraphics[width=0.85\columnwidth]{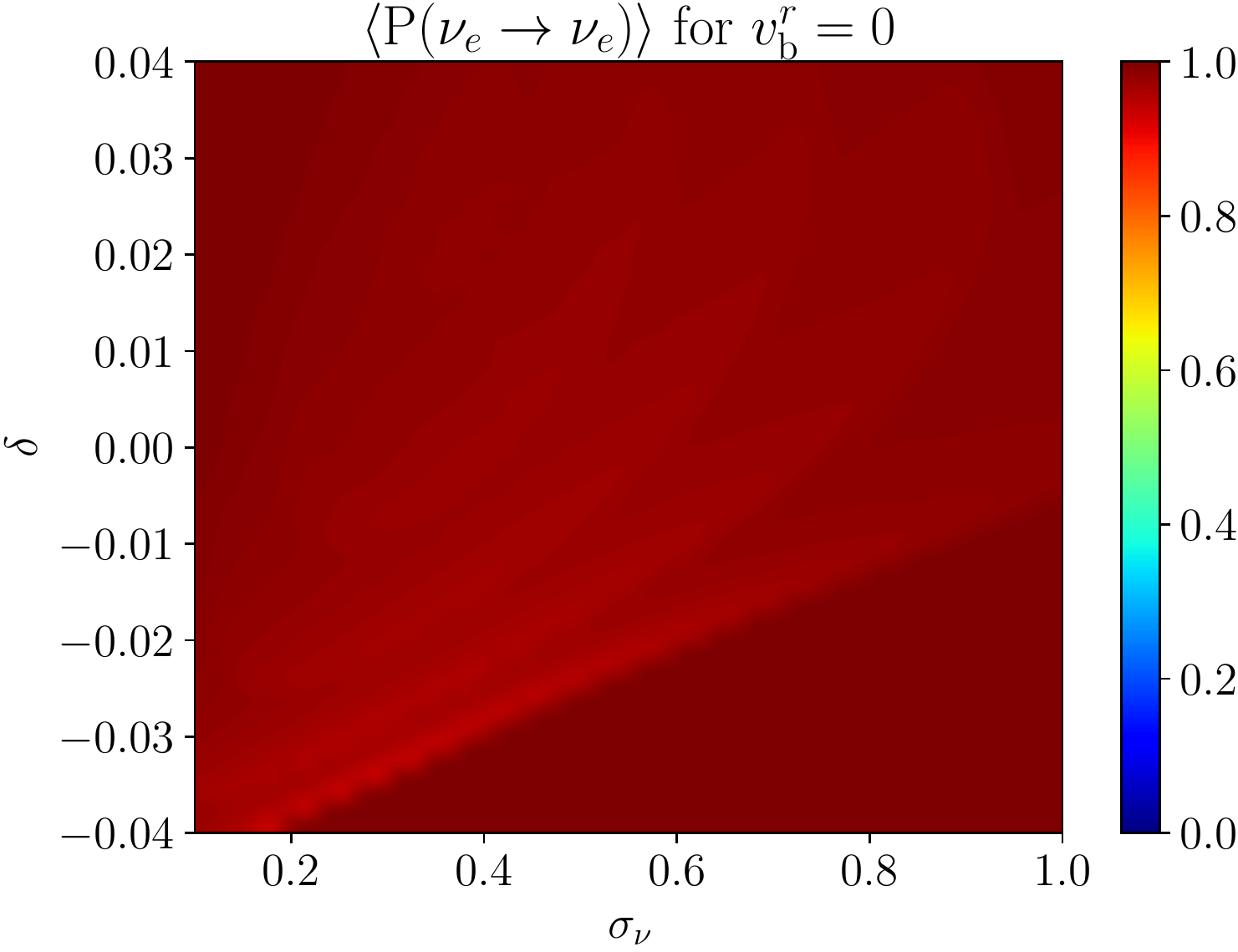}\hspace{1.0cm}  
\includegraphics[width=0.85\columnwidth]{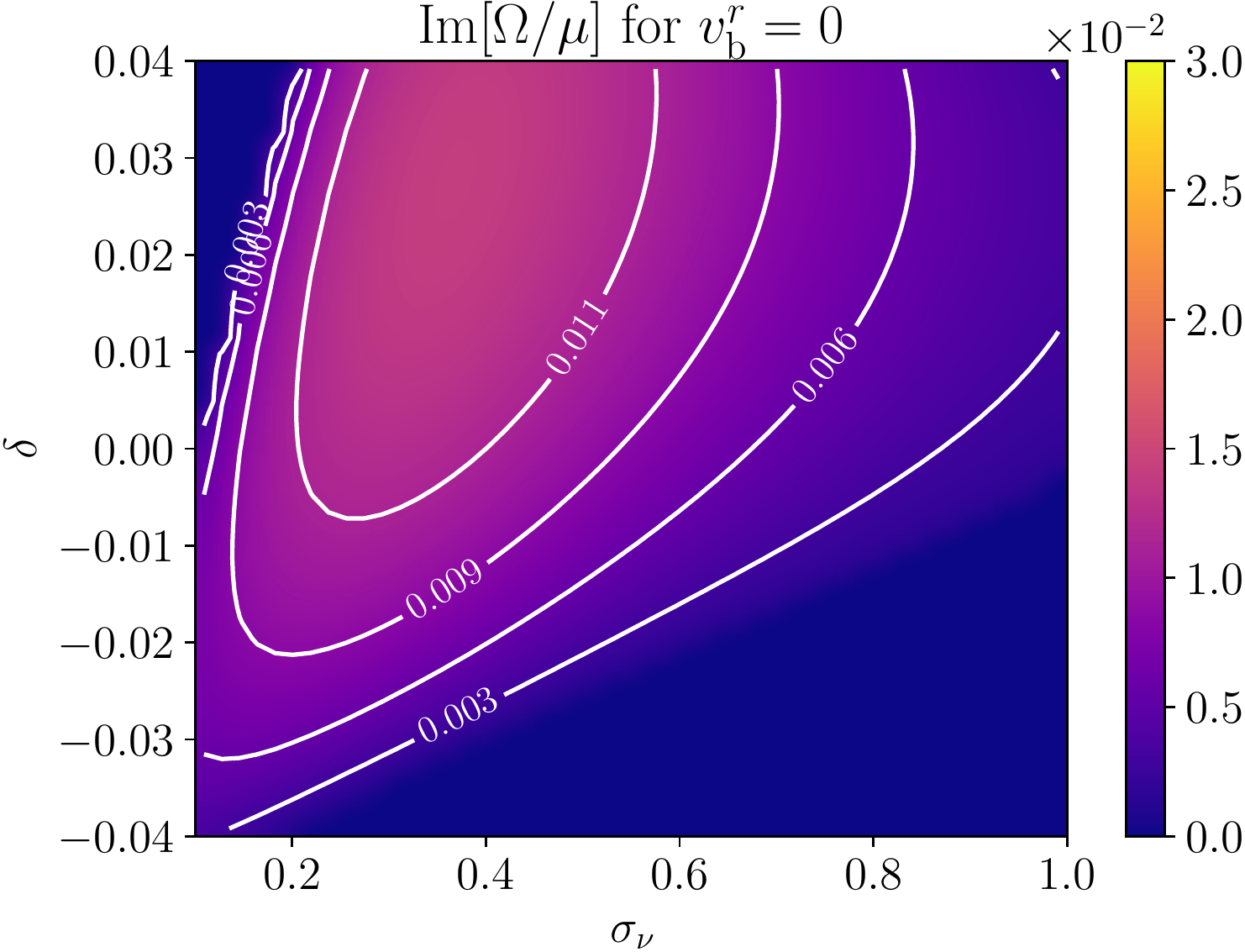}
\caption{ 
Isotropic case: Time averaged survival probabilities (left panel) and growth rates of the flavor instability (right panel) for the 2D parametric box. A large portion of the parameter space remains stable after $10^{-6}$ seconds, which is confirmed by the linear stability analysis (right panel). Moreover, for the parameter points that undergo flavor conversions, the final amount of non-electron flavor is minimal i.e. $\langle \mathrm{P}(\nu_e \rightarrow \nu_e) \rangle  \simeq 0.9$.
}
 \label{fig:5pr}
\end{figure*} 

\begin{figure*}[t!]
\centering
\includegraphics[width=0.85\columnwidth]{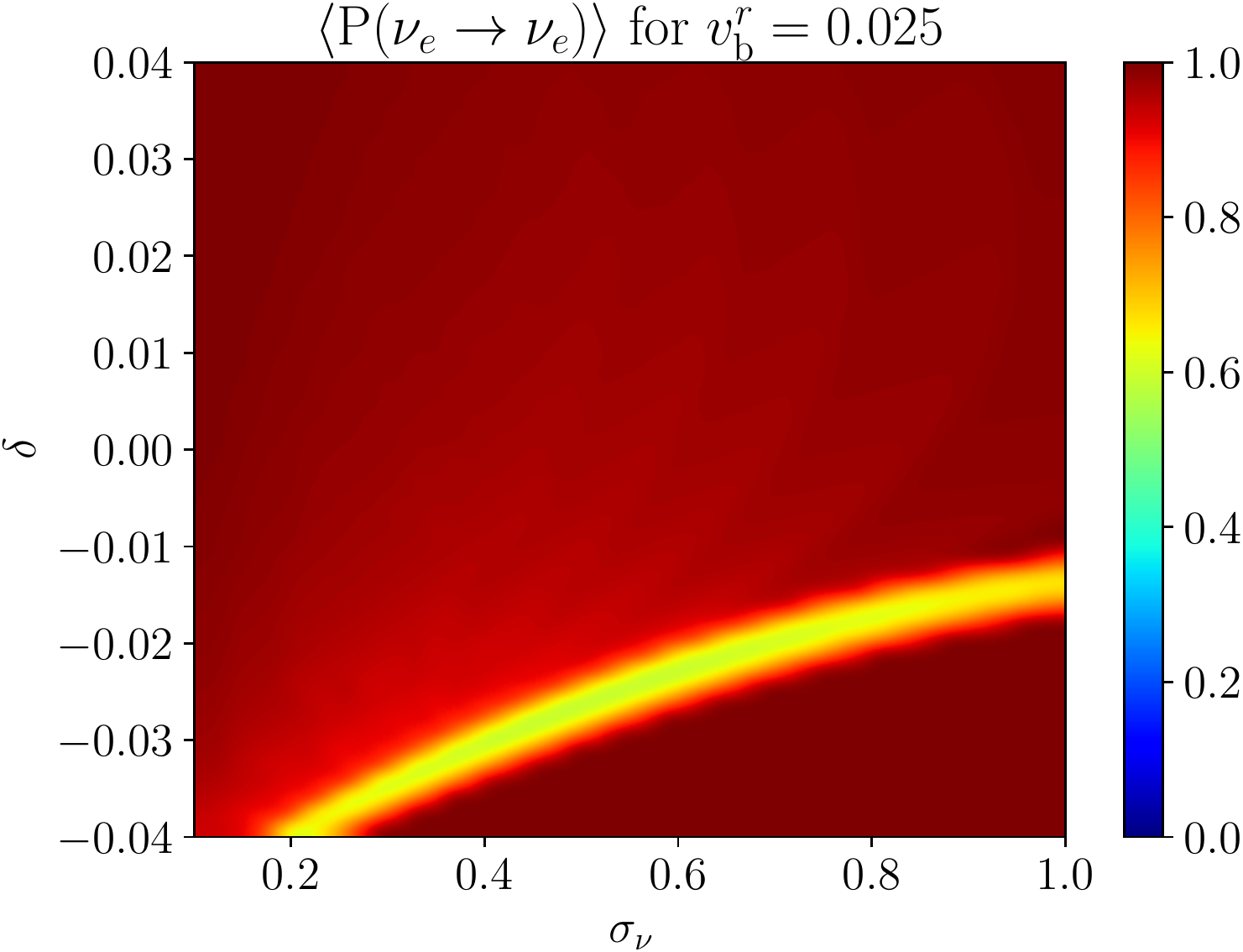}\hspace{1.0cm}
\includegraphics[width=0.85\columnwidth]{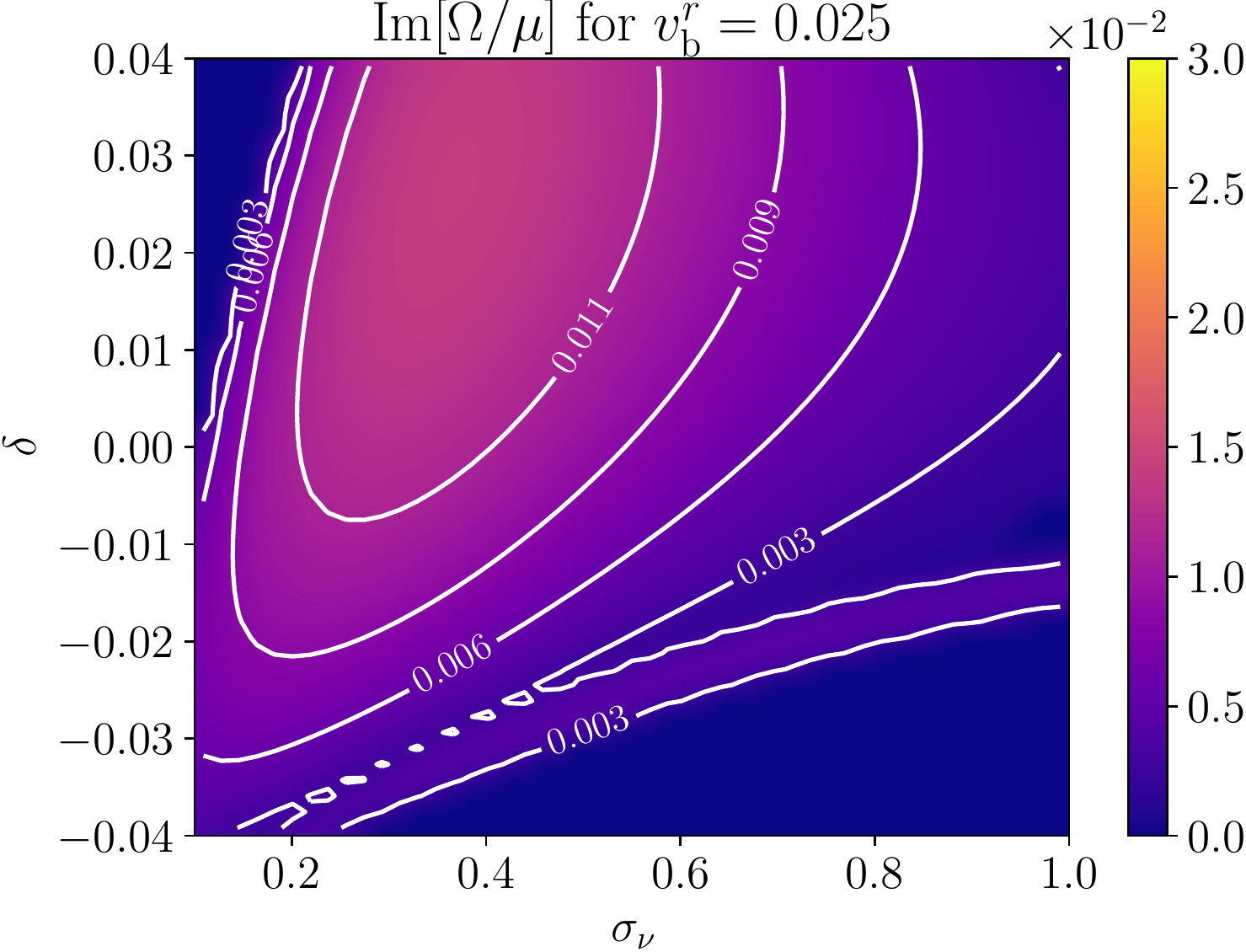}
\includegraphics[width=0.85\columnwidth]{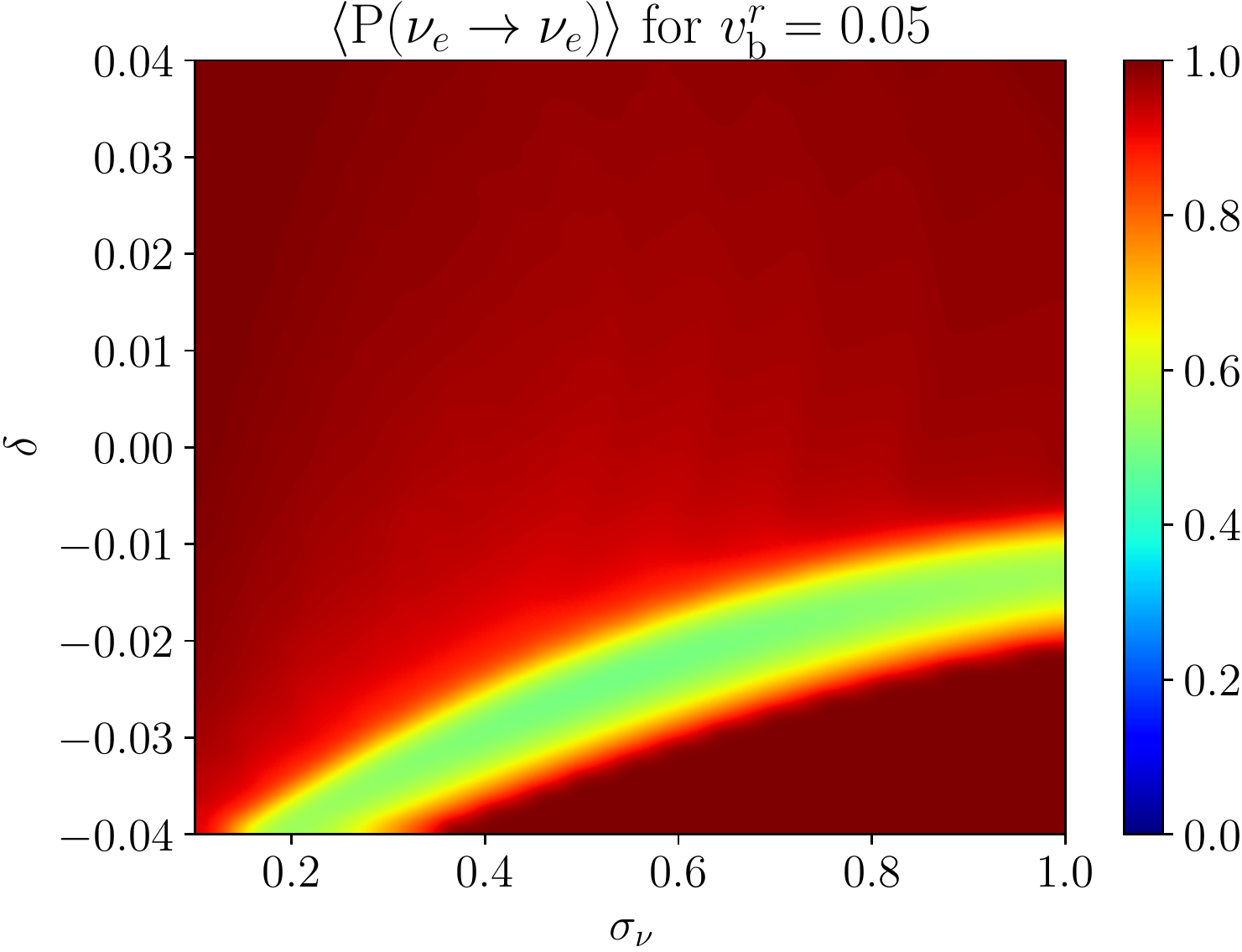}\hspace{1.0cm}
\includegraphics[width=0.85\columnwidth]{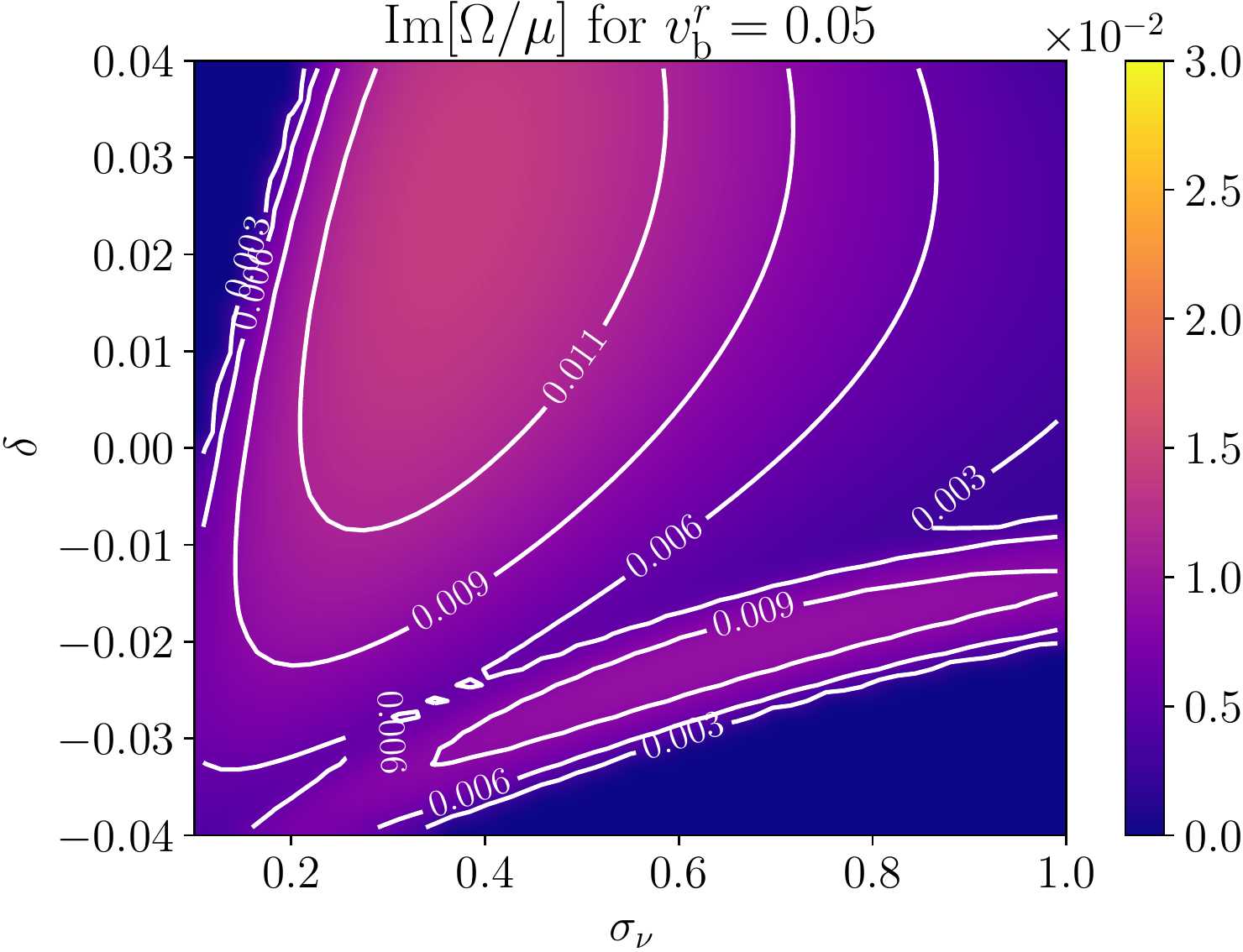}
\includegraphics[width=0.85\columnwidth]{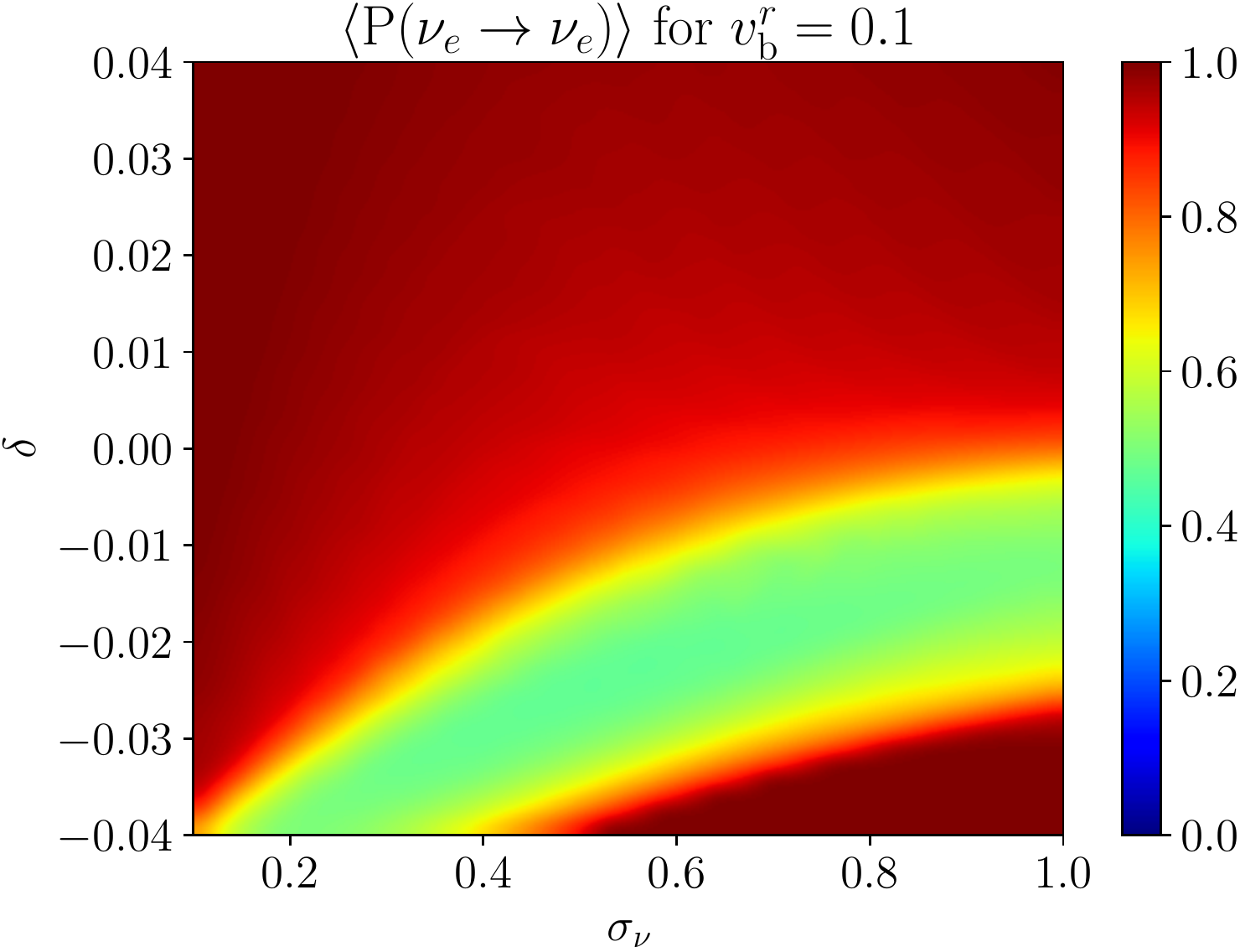}\hspace{1.0cm}
\includegraphics[width=0.85\columnwidth]{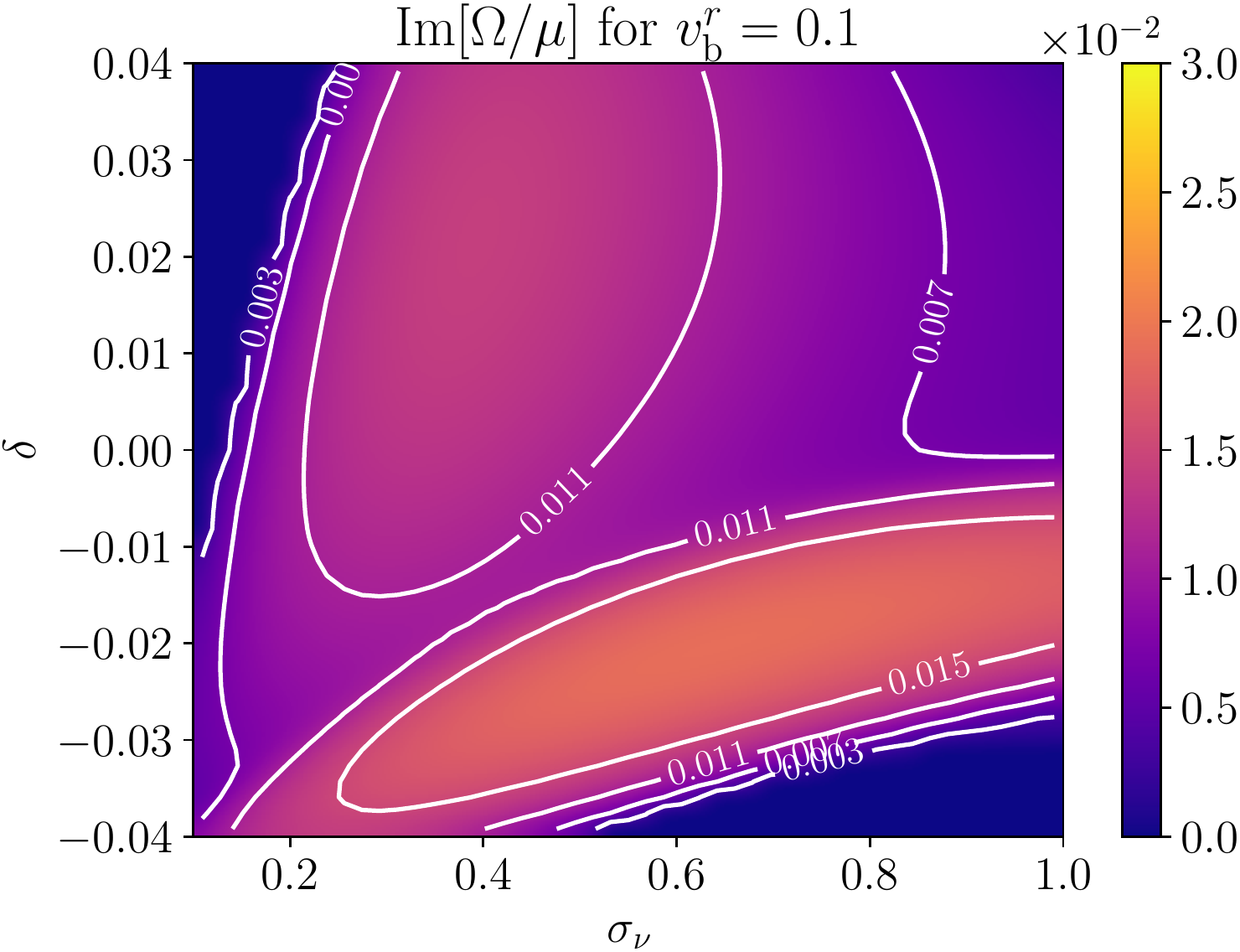}
\caption{ 
Time averaged survival probabilities (left panels) and growth rates of the flavor instability (right panels) for three selected values of the bulk velocity: $\vbr=0.025$ (top panels), $\vbr=0.05$ (middle panels) and $\vbr=0.1$ (bottom panels). As the value of the bulk velocity reaches $\vbr \simeq 0.025$ a new unstable region emerges in the bottom of the 2D parametric box where equipartition is achieved i.e. $\langle \mathrm{P}(\nu_e \rightarrow \nu_e)\rangle \simeq 0.5$. Moreover, not only the the conversions to the non-electron flavors are enhanced but also oscillations take place earlier due to larger growth rates, see left panels.
}
\label{fig:6pr}
\end{figure*}

\begin{figure*}[t!]
\centering
\includegraphics[width=0.85\columnwidth]{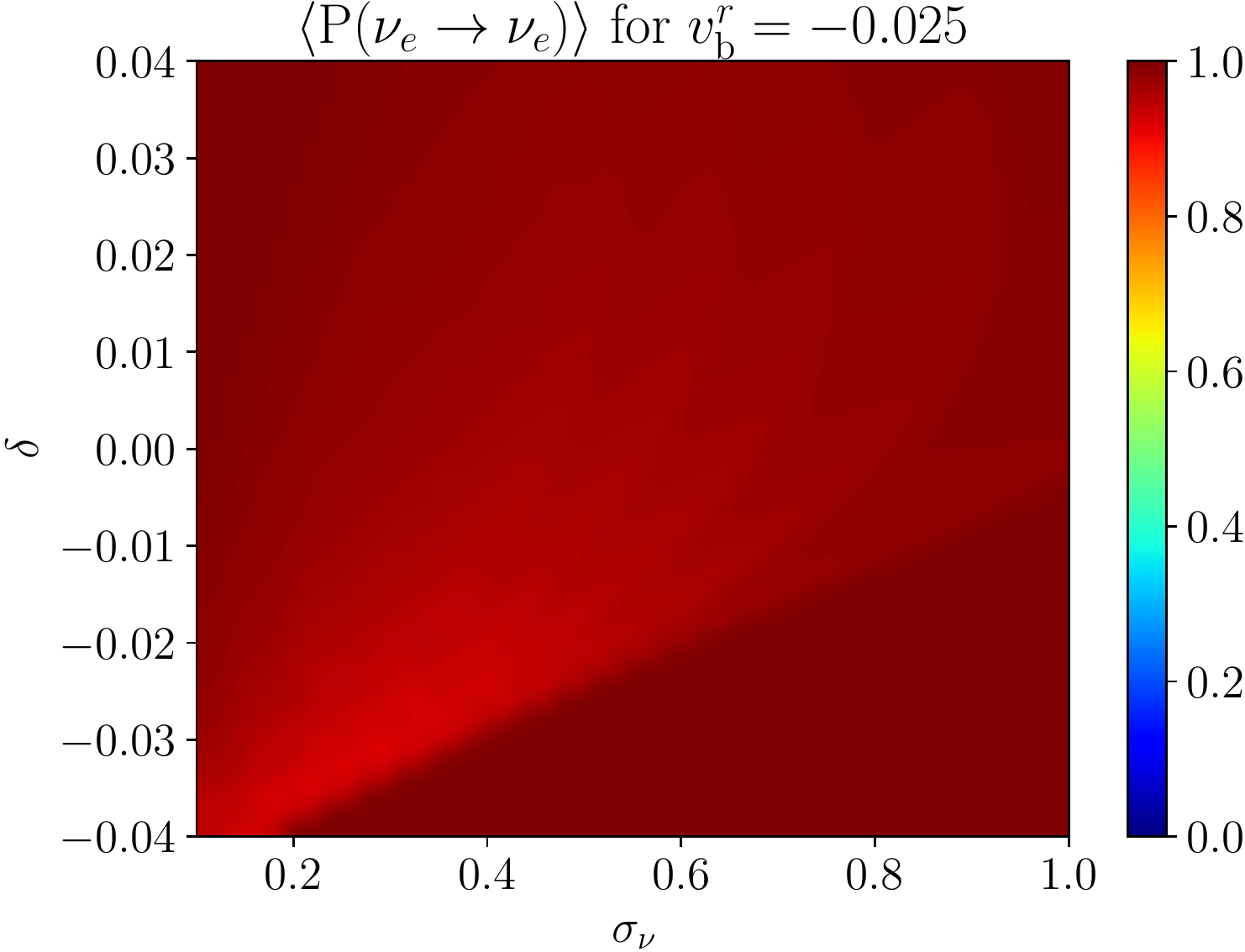}\hspace{1.0cm}  
\includegraphics[width=0.85\columnwidth]{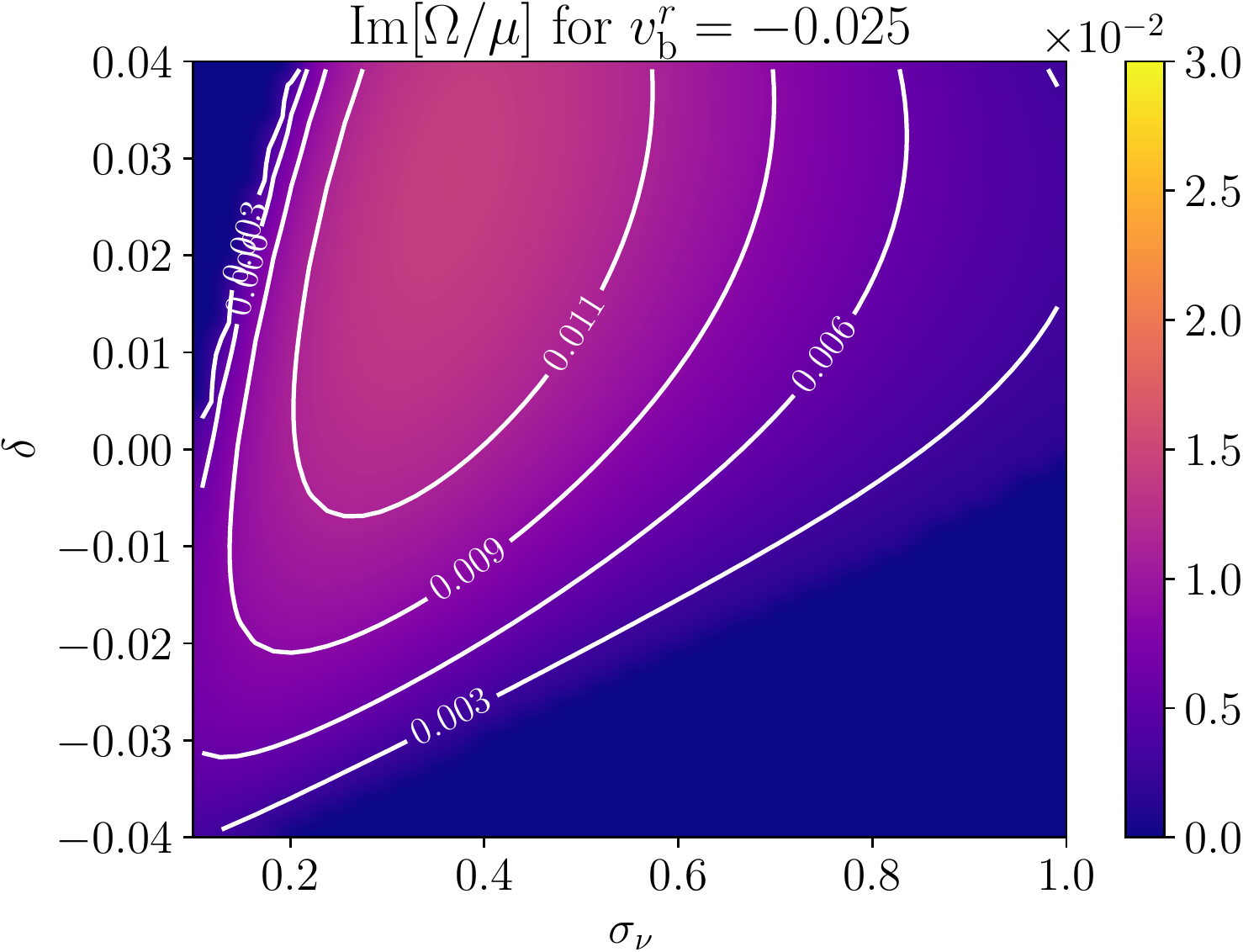}
\includegraphics[width=0.85\columnwidth]{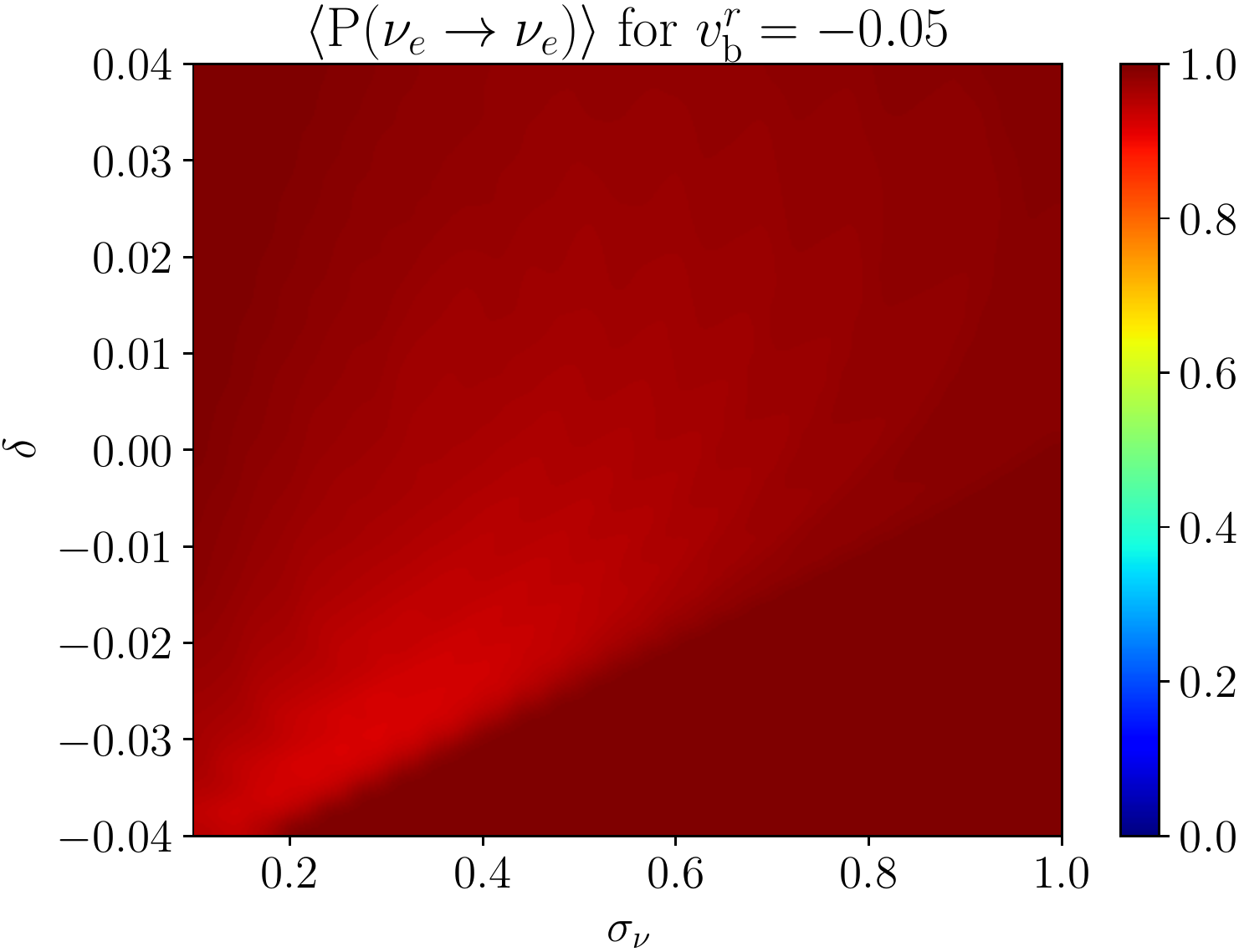}\hspace{1.0cm} 
\includegraphics[width=0.85\columnwidth]{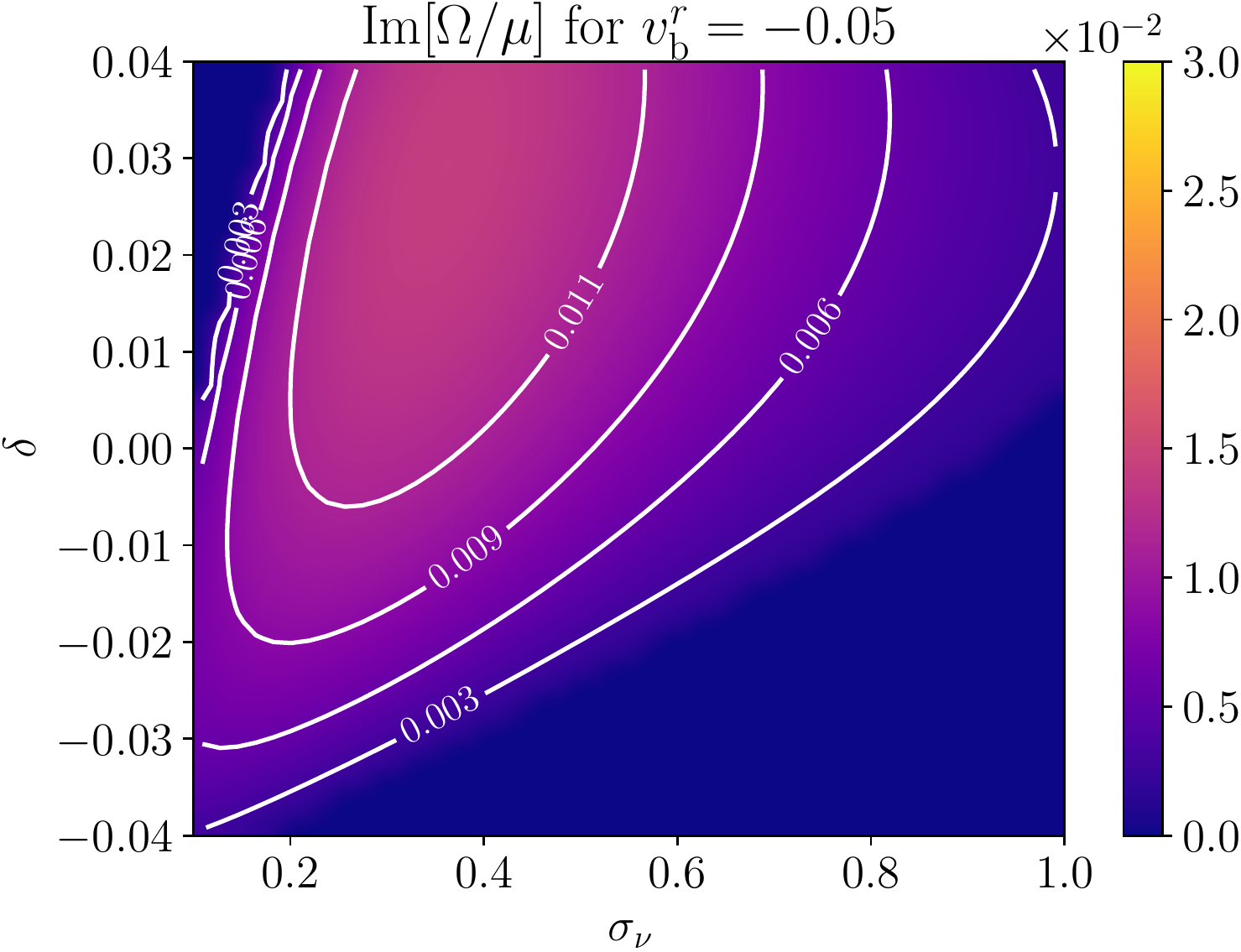}
\includegraphics[width=0.85\columnwidth]{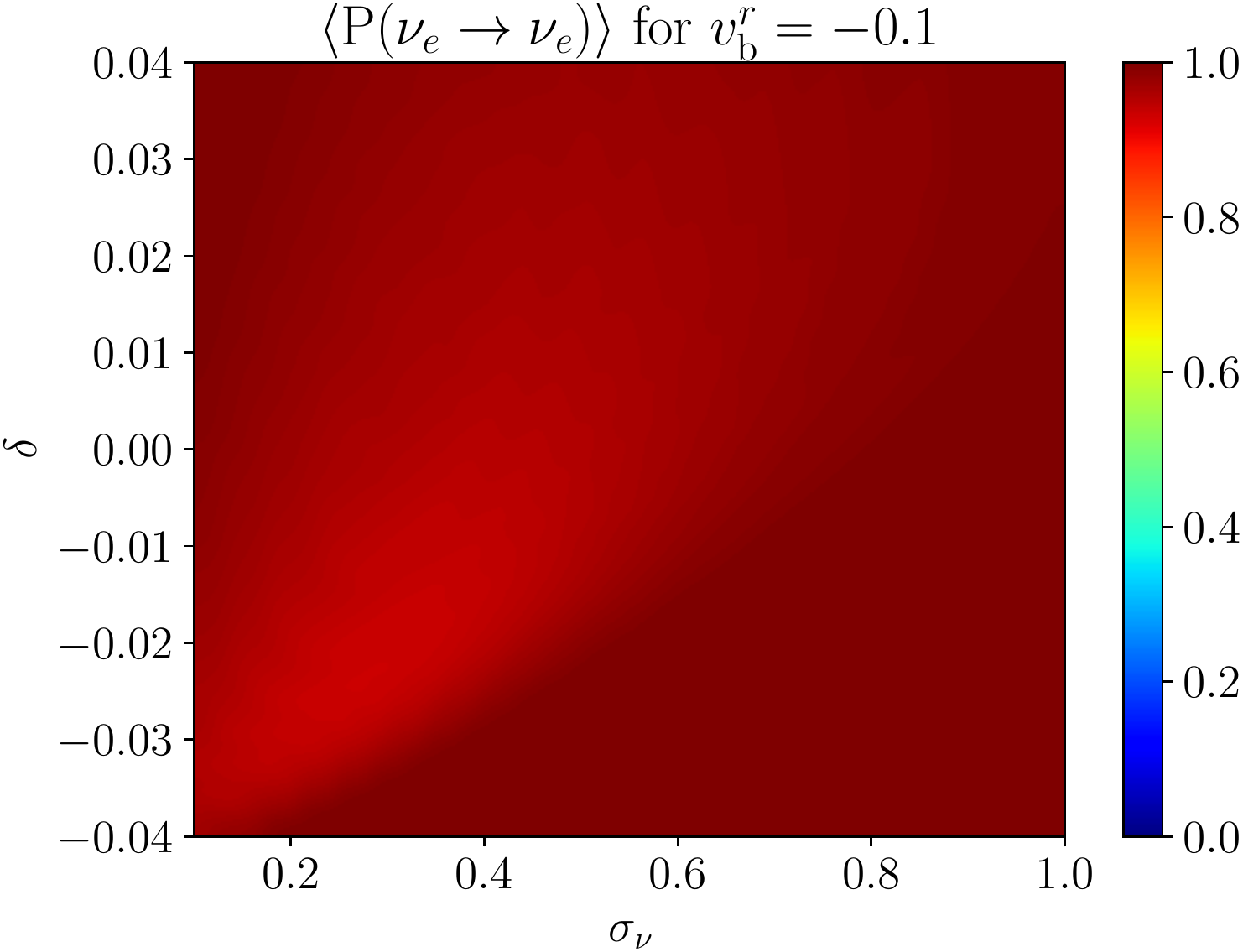}\hspace{1.0cm} 
\includegraphics[width=0.85\columnwidth]{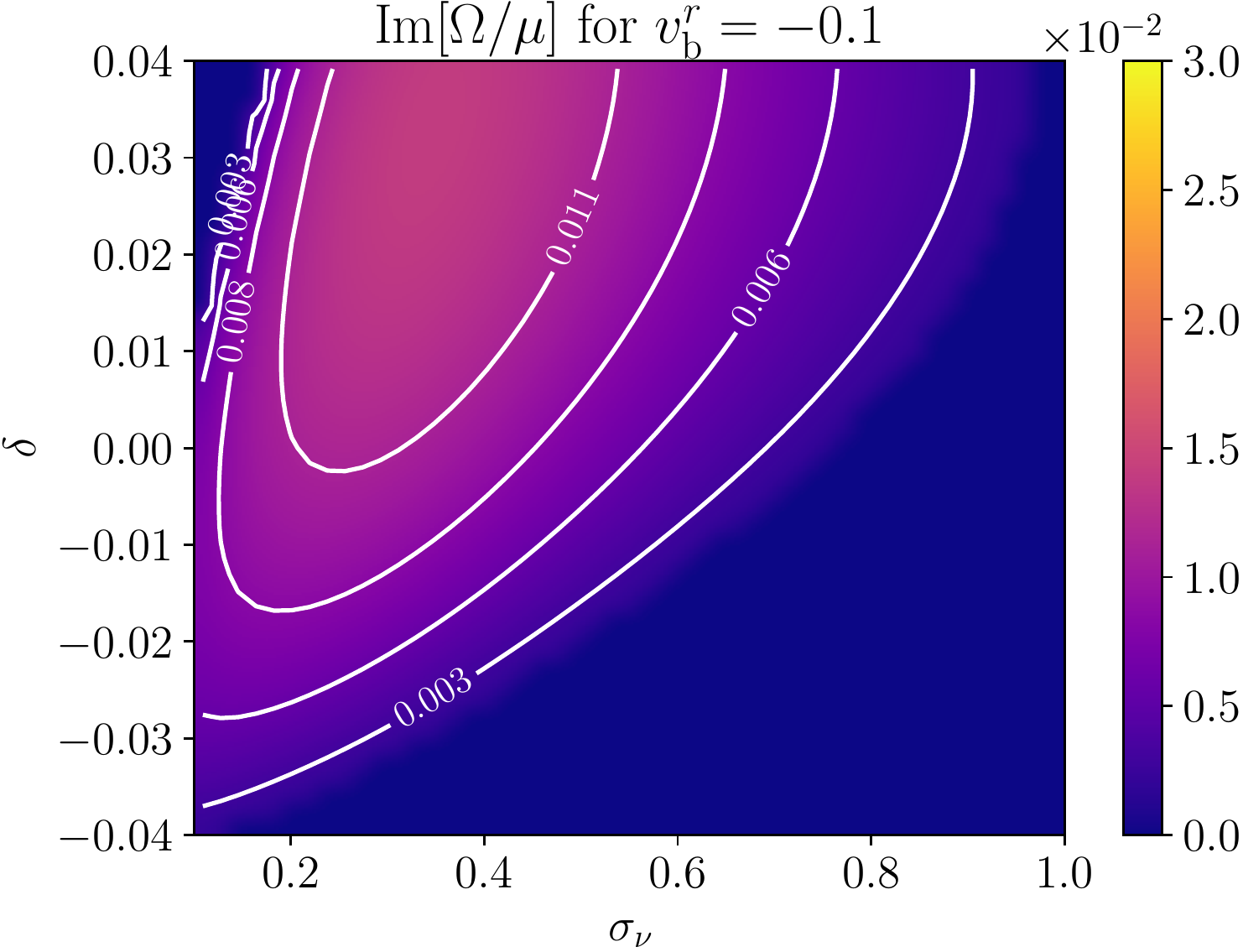}
\caption{ 
Time averaged survival probabilities (left panels) and growth rates of the flavor instability (right panels) for three selected values of the bulk velocity: $\vbr=-0.025$ (top panels), $\vbr=-0.05$ (middle panels) and $\vbr=-0.1$ (bottom panels). Qualitatively, the radially inward case is not very sensitive to the magnitude of $\vbr$ since the time averaged survival probabilities and the growth rates are very similar to each other. Interestingly, a radially inward matter suppresses oscillations and resembles, at least qualitatively, the isotropic case.
}
\label{fig:7pr}
\end{figure*} 

In Fig.~\ref{fig:4pr}, we show the dependence of the growth rate, $\mathrm{Im}(\Omega)$, on the bulk velocity for a representative case. For radially outward bulk velocity that is not large in magnitude, the effect of the bulk velocity is not significant. For a radially outward velocity in excess of $\sim 0.025$, one can see a dramatic change in the trend, and the growth rate of the flavor instability increases with the magnitude of the bulk velocity. This trend continues for magnitudes up to $\sim 0.3$ after which the trend is reversed. 
It should be noted that the neutrino angular distribution in Fig.~\ref{fig:4pr} is chosen to ensure that the dependence on the bulk velocity can be demonstrated. There are instances of angular distributions of neutrinos for which the effect of bulk velocity is not so dramatic. 

The presence of increased flavor instability as a result of the radially outward bulk velocity, as determined by the value of $\textrm{Im}(\Omega)$, should not necessarily imply an increased flavor conversion rate. However, as we show in the next section, the configurations for which the bulk velocity causes an increase in the flavor instability also result in an increased flavor conversion rate. This can be demonstrated by studying the effect of the bulk velocity on several different configurations of angular distributions as we do in the next section.


\section{2D parametric grid}
\label{sec:grid}

The results presented Fig.~\ref{fig:3pr} are very general for the angular distributions of the form presented in eq.~\ref{eq:init}. For all possible values of $\delta$ and $\sigma_{\nu}$, radially outward bulk velocity can enhance the neutrino flavor conversion rate, while radially inward bulk velocity always suppresses the neutrino flavor conversion rate. In order to demonstrate this, we calculate the growth rate of $|\rho_{ex}|$ in the linear regime with and without the inclusion of the bulk velocity. In the linear regime, the growth rate of the off-diagonal component $|\rho_{ex}|$ is exponential with time if the system is unstable, and we denote the growth rate by $\Omega$. 

We also calculate the survival probability of electron neutrinos. In some circumstances, the survival probability of neutrinos is periodic, and the time averaged survival probability is a much better representation of the amount of neutrino flavor conversion. We thus rely on the time averaged survival probability of neutrino, averaged over $t=10^{-6}$ seconds, which we denote by $\langle \textrm{P} (\nu_{e} \rightarrow \nu_{e}) \rangle $,
as a measure of the neutrino flavor conversion rate. The time averaged survival probability underestimates the flavor conversion rate in some instances due to the significant amount of time required to reach the nonlinear regime, but it is nonetheless a fair representation for the purpose of this paper.


\subsection{Isotropic $(\vbr=0)$}
 
In Figs.~\ref{fig:5pr} we show the flavor evolution of (anti)neutrinos for a wide range of values of $\sigma_{\nu}$ and $\delta$. Each point has a different neutrino angular distribution while the matter angular distribution is determined by $\vbr=0$ and kept unchanged. This provides a simple way of understanding the main differences between the isotropic and anisotropic scenarios for a wide range of (anti)neutrino angular distributions configurations and allows us to better understand which systems deviate more strongly from the isotropic scenario. In Fig.~\ref{fig:5pr} we present the time averaged survival probability $\langle \mathrm{P}(\nu_{e}\rightarrow \nu_{e}) \rangle$ across the grid. 

In the isotropic scenario, the only dimensionful quantity is $\mu$; neutrinos and antineutrinos oscillate in a bipolar fashion. This is explicitly shown in a special case in the top panels of Fig.~\ref{fig:3pr}, but this is true for all cases considered in Fig.~\ref{fig:5pr}.

	
\subsection{Radially outward $(\vbr > 0)$}

For comparison with the isotropic case, we perform the same numerical simulations as in Fig.~\ref{fig:5pr} but with $\vbr \neq 0$. In particular, each point on the grid has a matter angular distribution determined by $\vbr = 0.025, 0.05$ and $0.1$, all of which peak in the forward direction i.e., more electrons are emitted in the forward direction.

In Fig.~\ref{fig:6pr} we present results for the case where the matter has a positive velocity i.e., matter points along the forward direction. In a realistic supernova scenario, this would qualitatively correspond to the case where the post-shock material is pushed outwards and away from the PNS with a positive radial velocity.

First of all, because the flavor evolution now depends on two dimensionful quantities $\mu$ and $\lambda$, the oscillations are no longer (entirely) bipolar, and the interference caused by the matter term depends on the ratio $\lambda/\mu$. In particular, one can see that for some points on the grid, flavor conversions are unchanged, while in some others, the interference between $H_{e \nu}$ and $H_{\nu\nu}$ is very prominent. Secondly, a much larger region of the parameter space spanned by $\sigma_{\nu}$ and $\delta$ undergoes a significant amount of flavor conversions within the considered time window compared to the isotropic case. We observe an enhancement of oscillations in Fig.~\ref{fig:6pr} where $\langle \mathrm{P}(\nu_{e}\rightarrow \nu_{e})\rangle \simeq 0.5$ for a wider range of configurations. 

The panels on the right side of Fig.~\ref{fig:6pr} shows the growth rate of the off-diagonal components of the density matrix in the linear regime. The regions that show an enhanced flavor conversion probability, we also see a change in the growth rate as seen by the opening of the new regions of the parameter space for which the growth rate increases with increasing bulk velocity.


\subsection{Radially inward $(\vbr < 0)$}

For completeness, we investigate the other possible scenario where the radial velocity of the post-shock material is negative. In particular, each point on the grid has a matter angular distribution determined by $\vbr = -0.025, -0.05$ and $-0.1$, all of which peak in the backward direction.

In Fig.~\ref{fig:7pr} we show the same colormaps and angular distributions as in Fig.~\ref{fig:6pr} but with matter pointing in the backward direction while keeping the other parameters unchanged. Interestingly, we see the opposite trend as in the former forward-peaked case. Instead of leading to flavor equipartition, an isotropic backward matter potential leads a suppression of oscillations; even less neutrinos are converted in this case, see Fig.~\ref{fig:7pr} where $\langle \mathrm{P}(\nu_{e}\rightarrow \nu_{e}) \rangle \simeq 0.9$ at most. Similarly, in the right panels of Fig.~\ref{fig:7pr} we can see that the results of the linear stability analysis are also qualitatively unchanged in the case of bulk velocity that is radially inwards.


\section{Conclusions}
\label{sec:conclusion}

We show that since the fast flavor evolution of neutrino depends strongly on the angle dependence of the coherent forward scattering potential, the angle dependence of the matter potential cannot be ignored. The angle dependence in the matter potential can naturally arise due to the bulk velocity of matter in astrophysical environments. In some instances, the bulk velocity of matter present in astrophysical environments can be as large as $10\%$ of the speed of light. We find that bulk velocities much smaller than the maximum allowed velocities can substantially change the neutrino flavor conversion rate. 

We obtain quantitative estimates in this paper to support our understanding. There are two alternative ways of understanding the reason for the impact of the bulk velocity of matter on neutrino flavor evolution. One, on which we heavily rely in this paper, is to consider the problem in terms of the modification in the angular dependence of the potential experienced by the neutrinos due to the bulk velocity of matter. The neutrinos traveling in the direction of the bulk velocity of the matter will see a smaller flux compared to the neutrinos in the opposite direction giving rise to the angle dependence of the matter potential.
Another way to think about the problem is to consider the problem in the rest frame of the matter. Due to the change in the reference frame, the angular distribution of neutrinos is modified and could be modified in a way that either enhances or suppressed the neutrino flavor conversion rates. It should be noted that the two approaches mentioned here are two different ways of think about the same phenomenon; in two different reference frames.

Notwithstanding, the effect of the bulk velocity on the neutrino flavor evolution is far from negligible, the necessary condition for the existence of fast flavor instability, the presence of ELN crossings, remains unchanged. We have explored several cases of neutrino angular distribution without an ELN crossing, to examine whether the anisotropic matter term can lead to fast flavor instability. However, we were not able to find a case in which fast flavor conversions were present in the absence of ELN crossings irrespective of the angle dependence of the matter term. This finding is not at all surprising since it is possible to go to a reference frame in which the matter term is isotropic and the ELN crossings are features that are not dependent on the reference frame. 

Although in this paper, we only consider two extreme possibilities of the bulk of matter that is either radially outward or inward, this may not be the case in a realistic astrophysical system. However, we have restricted our analysis to these possibilities, to not clutter the analysis with too many variables. The possible enhancement of neutrino flavor conversion rates due to the bulk velocity of matter can be clearly demonstrated in the setup that is considered in this paper.

The demonstration of possible enhancement of neutrino flavor conversion rate due to the bulk velocity of matter also raises several important issues that are of relevance in the supernova mechanism. In the widely popular delayed neutrino driven supernova mechanism, the emphasis is on investigating the role of neutrinos in triggering convection; however, the role of the convective flow of matter in triggering neutrino flavor evolution can possibly lead to a feedback mechanism that is either positive or negative. Future analysis of the subject matter with realistic velocity profiles may be able to shed more light on the matter. 


\acknowledgments 
IPG would like to thank Irene Tamborra for the support during the course of the project. SS is supported by the Villum Foundation (Project No. 13164). IPG acknowledges support from the Danmarks Frie Forskningsfonds (ProjectNo. 8049-00038B).



\bibliography{references.bib}

\appendix

\end{document}